\documentclass[twocolumn]{aastex631}

\newcommand{\jwst}{{JWST}}
\newcommand{\lenstool}{\texttt{Lenstool}}

\newcommand{\lstar}{L${_*}$ Gal.}

\newcommand{\nconstraints}{275}
\newcommand{\nsysnew}{13~}
\newcommand{\finalrms}{$0\farcs23$}

\usepackage{hyperref}
\usepackage{amsmath}
\usepackage{soul}
\usepackage[normalem]{ulem}
\usepackage{color}

\defcitealias{lagattuta2023}{L23}
\defcitealias{lagattuta2026}{L26}

\begin{document}

\title{Resolving 3 Exotic Hyperbolic-Umbilic Lensing Configurations in the `Cosmic Mantis': \\ An Exploration of RXJ0437.1+0043 with JWST}

\author[0000-0002-8261-9098]{Catherine Cerny}
\affiliation{Department of Astronomy, University of Michigan 
1085 South University Avenue 
Ann Arbor, MI 48109, USA}
\email{cecerny@umich.edu}

\author[0000-0002-7633-2883]{David J. Lagattuta}
\affiliation{Centre for Astrophysics Research, Department of Physics, Astronomy and Mathematics, University of Hertfordshire, Hatfield AL10 9AB, UK}
\affiliation{Centre for Extragalactic Astronomy, Department of Physics, Durham University, South Road, Durham DH1 3LE, UK}
\affiliation{Institute for Computational Cosmology, Durham University, South Road, Durham DH1 3LE, UK}

\author[0000-0002-7559-0864]{Keren Sharon}
\affiliation{Department of Astronomy, University of Michigan 
1085 South University Avenue 
Ann Arbor, MI 48109, USA}

\author[0000-0002-3475-7648]{Gourav Khullar}
\affiliation{Department of Astronomy, University of Washington, Physics-Astronomy Building, Box 351580, Seattle, WA 98195-1700, USA}

\author[0000-0002-0443-6018]{Benjamin Beauchesne}
\affiliation{Centre for Extragalactic Astronomy, Department of Physics, Durham University, South Road, Durham DH1 3LE, UK}
\affiliation{Institute for Computational Cosmology, Durham University, South Road, Durham DH1 3LE, UK}

\author[0000-0002-3398-6916]{Alastair C. Edge}
\affiliation{Centre for Extragalactic Astronomy, Department of Physics, Durham University, South Road, Durham DH1 3LE, UK}
\affiliation{Institute for Computational Cosmology, Durham University, South Road, Durham DH1 3LE, UK}

\author[0000-0003-1974-8732]{Mathilde Jauzac}
\affiliation{Univ Toulouse, CNES, CNRS, IRAP, Toulouse, France}
\affiliation{Centre for Extragalactic Astronomy, Department of Physics, Durham University, South Road, Durham DH1 3LE, UK}
\affiliation{Institute for Computational Cosmology, Durham University, South Road, Durham DH1 3LE, UK}
\affiliation{Astrophysics Research Centre, University of KwaZulu-Natal, Westville Campus, Durban 4041, South Africa}
\affiliation{School of Mathematics, Statistics \& Computer Science, University of KwaZulu-Natal, Westville Campus, Durban 4041, South Africa}

\author[0000-0001-6636-4999]{Marceau Limousin}
\affiliation{Aix Marseille Univ, CNRS, CNES, LAM, Marseille, France }

\author[0000-0003-3266-2001]{Guillaume Mahler}
\affiliation{STAR Institute, Quartier Agora - All\'ee du six Ao\^ut, 19c B-4000 Li\`ege, Belgium}

\author[0000-0001-6278-032X]{Lukas J. Furtak}
\affiliation{ Department of Astronomy, The University of Texas at Austin, Austin, TX, 78712, USA }
\affiliation{ Cosmic Frontier Center, The University of Texas at Austin, Austin, TX 78712, USA}

\author[0000-0002-5899-3936]{Leo Fung}
\affiliation{Centre for Extragalactic Astronomy, Department of Physics, Durham University, South Road, Durham DH1 3LE, UK}
\affiliation{Institute for Computational Cosmology, Durham University, South Road, Durham DH1 3LE, UK}

\author[0000-0003-3672-9365]{Qiuhan He}
\affiliation{Kapteyn Astronomical Institute, University of Groningen, Groningen, The Netherlands}

\author[0000-0002-6085-3780]{Richard J. Massey}
\affiliation{Centre for Extragalactic Astronomy, Department of Physics, Durham University, South Road, Durham DH1 3LE, UK}
\affiliation{Institute for Computational Cosmology, Durham University, South Road, Durham DH1 3LE, UK}

\author[0000-0002-7876-4321]{Ashish K. Meena}
\affiliation{Department of Physics, Indian Institute of Science, Bengaluru 560012, India}

\author[0000-0002-0350-4488]{Adi Zitrin}
\affiliation{Department of Physics, Ben-Gurion University of the Negev, PO Box 653, Be’er-Sheva 8410501, Israel}


\begin{abstract}
We present an updated strong lensing model for the galaxy cluster RXJ0437.1+0043 (z=0.285) using new JWST NIRCam imaging in four pass-bands. The cluster field notably displays at least three separate gravitationally lensed galaxies in hyperbolic-umbilic (H-U) configurations, which in the JWST imaging show exquisitely resolved stellar clumps at scales as small as $0\farcs04$. We identify 13 new systems of multiply-imaged lensed galaxies and determine the redshifts for two sources based on archival spectroscopic observations. We also identify over 50 individually resolved substructure clumps in the multiple images of eight galaxies and add them into the lens model, for a total of \nconstraints~positional lensing constraints between all 26 lensed sources used in the model. We investigate the impact of different modeling choices on our results. In the final lens model, we reproduce the observed multiple images with a precision of \finalrms, and achieve $0\farcs13$ specifically for the H-U systems. We estimate a total enclosed mass in the cluster's strong lensing regime of $M(<110\mathrm{ kpc})\sim8.2\times10^{13}M\odot$, reconstruct the source images of the most highly magnified galaxies, and measure time delays in the most magnified, morphologically detailed systems. The isotropic lensing effect from the H-U configurations enables future detailed studies of the properties of their source galaxies, and the number and spatial density of additional constraints improve the model's ability to detect dark matter substructures within the cluster core. 
\end{abstract}




\section{Introduction} \label{sec:intro}

 The nature of dark matter (DM) is a tantalizing question in astrophysics. While its presence has been indirectly detected at a variety of scales \citep{vegetti2014,umetsu2018}, the largest quantities of DM are attributed to clusters of galaxies, which are the most massive gravitationally-bound objects in the Universe. The amount and distribution of DM in clusters can be determined through many techniques, such as gravitational lensing, which distorts and magnifies the intrinsic appearances of background galaxies as their emitted light travels through the plane of a massive lens between the background galaxy and the observer \citep[e.g., for a recent review, see][]{natarajan2024}. Lensing is uniquely suited to DM studies because it does not require \textit{a-priori} assumptions about physical properties of the cluster that may be difficult to measure (e.g., the 3-D structure of the cluster; its relaxation state; hydrostatic equilibrium; and scaling relations; \citealt{bartelmann2010}). In this work, we specifically use strong lensing (SL), which creates multiple highly-magnified images of background sources, to model the total mass of a galaxy cluster. SL models broadly rely on the number and spatial distribution of images of lensed sources that can be identified in the cluster field (e.g., \citealt{johnsonsharon2016}), as these images serve as direct constraints on the mass of the cluster.
 
 Lensing configurations that produce images close to the center of the cluster (e.g., within $\sim50$ kpc) are particularly valuable when investigating cosmological questions surrounding $\Lambda$CDM since they provide tight constraints on the inner slope of the mass density profile of the cluster \citep{sand2002, harvey2019,robertson2019,cerny2025,beauchesne2025}. One notable configuration is hyperbolic-umbilic (H-U) systems, which are a type of `exotic' lens \citep[e.g.,][]{petters2001, orban2009,meena2020} characterized by extremely magnified ($\mu>100$) images located within the cluster core. H-Us produce high numbers of multiple images with very low shape distortions, which makes them uniquely sensitive to the presence of small-scale mass components of galaxy clusters, such as dark matter subhalos \citep{natarajan2017,meena2023,doppel2026,lagattuta2026}. While H-Us have historically been quite rare (Abell 1703 was the only known example in literature for over a decade; \citealt{limousin2008}), recent observations with the Multi-Unit Spectroscopic Explorer (MUSE; \citealt{bacon2010}) and JWST have enabled the detection of new H-U systems, increasing the current known number to over 25 (D. Lagattuta in prep.). 

 JWST's ability to reveal lensed background sources has led to a new era of high-precision strong lens models. The wavelength coverage and sensitivity of JWST NIRCam imaging is well-suited to both discovering sources that are too red or faint to appear in HST or ground-based imaging, and for resolving stellar `clumps' as small as $0\farcs04$ within these lensed galaxies \citep{furtak2023,bergamini2023,mahler2023, bradley2025,cernyslice}. Resolving these clumps yields insights into the star formation history and evolution of their host galaxies \citep{vanzella2022,Claeyssens23,Claeyssens24,meena2023,vanzella2024,messa2025,hutchison2025,khullar2026}. In lensed galaxies, these substructure clumps can also be used as constraints in strong-lensing models, increasing the resolution of the lens model in regions with many constraints \citep{diego2023, bergamini2023, mahler2023, furtak2024, rihtar2026, limousin2026, abedi2026}.

 We present an updated lens model of the galaxy cluster RX\,J0437.1$+$0043 (hereafter RXJ0437; $z=0.285$; R.A. $=69.2896756$, Decl. $=0.7311474$) using newly-acquired deep JWST imaging. This cluster is uniquely suited to examining dark matter because it contains at least three distinct H-U lensed galaxies at three different redshifts, which were first identified in \citet[][hereafter L23]{lagattuta2023} using HST imaging and VLT/MUSE spectroscopy. We dub this cluster the `Cosmic Mantis' due to the presence of a very highly magnified galaxy that lies in between one of these H-U systems, creating an image of `eyes' and `arms' that looks strikingly similar to a mantis. \citetalias{lagattuta2023} created the first lens model for this cluster and found that it possesses a fairly elongated elliptical mass distribution (ellipticity $\sim0.4$), which forces the two primary critical curves (in the tangential and radial directions) to nearly intersect. The close proximity of the tangential and radial curves creates an environment where exotic lensing configurations can be generated \citep{basto2026}. In this cluster, H-U systems appear whenever these critical curves nearly touch (\citetalias{lagattuta2023}). 
 
 We use JWST NIRCam observations in four bands from Cycle 3 GO-6207 to identify an additional 13 systems of multiply-imaged galaxies in RXJ0437. We also identify and map a total of 57 emission clumps between the multiple images of eight galaxies. The total number of positional lensing constraints  (hereafter referred to as ``arcs'') used in the model is  \nconstraints. These identifications represent nearly a factor of three increase in the total number of arcs compared to the previous model, and also place much tighter constraints on the lensing potential around the H-U systems due to the higher number of clumps in each image, where the number of clumps increases, for example, from 7 clumps to 24 clumps in a single image of an H-U galaxy.

 In this paper, we investigate the effects of these additional arcs on the lensing model's ability to map the dark matter components of the cluster, from the cluster-scale DM halo to the individual substructures identified in \citetalias{lagattuta2023} and \cite{lagattuta2026}. We also examine some of the source properties of the most highly-magnified galaxies in the field.

This paper is organized as follows. Section~\ref{sec:data} describes the JWST imaging and archival spectroscopy. Section~\ref{sec:lensmodel} contains details on the updated strong lens model for RXJ0437. Section~\ref{sec:results} presents the primary findings from the JWST imaging and the new strong lens model. Section~\ref{sec:discussion} evaluates the impact of the inclusion of the clumps on the model and discusses the characteristic properties of the lensed galaxies. Section~\ref{sec:conclusions} describes the results and highlights future work that will be enabled with this model. We assume a standard $\Lambda$CDM cosmology with $\Omega_M~=~0.3$, $\Omega_{\Lambda}~=~0.7$, and $H_0$~=~70 km s$^{-1}$ Mpc$^{-1}$. All magnitudes are given in the AB system.

\section{Data} \label{sec:data}

\subsection{JWST Imaging}

JWST/NIRCam imaging of RXJ0437 took place as part of JWST Cycle 3 GO-6207 (PI: Lagattuta) on 2025 Jan 24 in four wide filters in the short and long wavelength channels: F090W (8417 s), F200W (4123 s), F277W (4123 s), and F444W (8417 s). We used the Shallow 4 mode, with only Module B active to stay within data allocation limits. Uniform coverage of the cluster is obtained through an IntramoduleBox 4 dither pattern, with an additional 4 sub-pixel Small-Grid-Dither dithers to fully sample the PSF at the shortest wavelengths in the short and long channels. The exposure time is motivated by the science goal, detection of $\leq 10^9 M_\odot$ substrucutres in the H-U region, which requires a per-pixel SNR $\sim 100$ \citep{he2023}. 

The data reduction for JWST imaging was conducted via a custom reduction pipeline (see \citealt{Rigby2023, cernyslice}), modified from the STScI \texttt{jwst} pipeline v1.18.0 and CRDS calibration reference data system pipeline mapping pmap 1364. We use Level 1b products from MAST using an STScI-adapted custom Python script. We process Level 2A data products with the built-in ``flicker noise" correction code \citep{rauscher2024} to correct for 1/f noise and jumps between amplifiers. We then use these processed Level 2b products to generate Level 3 dithered science-ready mosaics. The filters are WCS-matched to Gaia as per the specifications of the default pipeline.

\subsection{VLT/MUSE Spectroscopy}
We make use of spectroscopy from two VLT/MUSE observations, which we briefly summarize here. These observations consist of a 1 hour exposure taken on 2020 February 15 as a part of the Kaleidoscope Cluster Survey (PID 0104.A0801; PI A. Edge; \citealt{cerny2025}), obtained in WFM-NOAO-N mode; and a combination of eight deeper MUSE pointings observed between January 15 and February 13 2021 (PID 106.21 AD; PI D. Lagattuta), which were acquired using Adaptive Optics corrections in WFM-AO-N mode. We refer the reader to \citetalias{lagattuta2023} and \cite{lagattuta2026} for further information regarding the reduction and combination of these observations.

\begin{figure*}
 \centering
\includegraphics[width=0.90\linewidth]{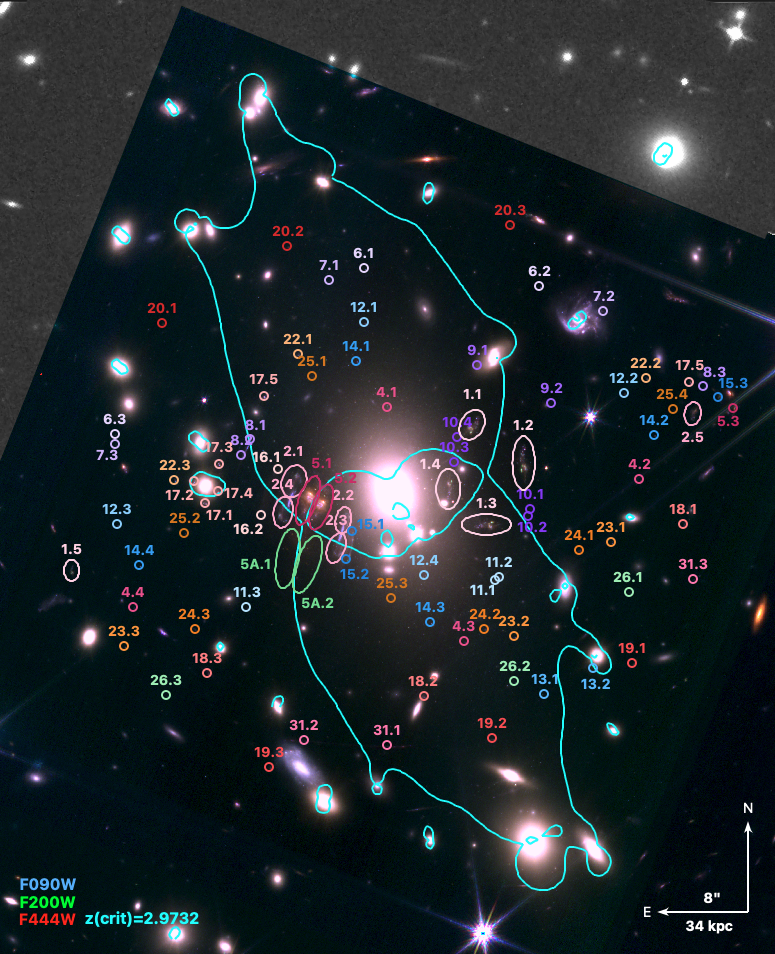}
    \caption{False color image of RXJ0437, constructed using JWST NIRCam imaging with F444W in the red channel, F200W in the green, and F090W in the blue. Archival HST/F110W imaging is shown in grayscale at the edge of the field to display the location of the third brightest cluster member galaxy. The critical curves for the fiducial model are shown in cyan at $z=2.9732$, which is the redshift of the H-U labeled Source \#1. The multiple images of lensed galaxies that were used as constraints in the model are marked in different colored circles, where each color corresponds to a single system of lensed sources. The multiple images are labeled as \#X.\#Y, where \#X corresponds to a multiply-imaged family (or system), and \#Y is a specific image from that family. Clumps used in each source are not shown in this figure for visual clarity; the clumps in systems \#1, \#2, and \#5 are indicated in \autoref{fig:HUsnapshots}.
    }\label{fig.clustermodel}
\end{figure*}

\section{Lens Model} \label{sec:lensmodel}
The \jwst~imaging reveals several new lensing systems, as well as numerous luminous ``clumps'' in the brightest lensed galaxies. We use this information to update and improve upon the HST-based lens model presented in \citetalias{lagattuta2023}, using the parametric MCMC modeling algorithm \lenstool~\citep{jullo07}. 
We model RXJ0437 (i.e., the lens) as a set of DM halos, using cluster-scale components for large, ``smooth'' potentials and smaller galaxy-scale halos to account for individual cluster members. We describe all components as truncated pseudo-isothermal elliptical (dPIE; \citealt{eliasdottir2007}) mass halos. Halo parameters are optimized in the model, using the positions of multiply-imaged galaxies and clumps within them as constraints.

The halos are described with seven parameters: the mass centroid ($x, y$); ellipticity and position angle ($\epsilon, \theta$); characteristic radii that describe the points at which the profile deviates from a simple isothermal profile ($r_{\mathrm{core}}, r_{\mathrm{cut}}$); and a normalization ($\sigma_0$) that serves as a scaled version of the halo velocity dispersion. All the parameters for the cluster-scale DM halo in this cluster are optimized within the model, with the exception of the truncation radius $r_{\mathrm{cut}}$, which is fixed to $1500$ kpc. This is because the cut radius typically extends well beyond the region of the cluster containing multiply-imaged galaxies \citep{limousin2007}, making it difficult to constrain with strong lensing alone.

The cluster members are selected by their color relative to the cluster red sequence \citep{gladdersyee2000}, guided by the locus of spectroscopically-confirmed cluster member galaxies in color-magnitude space (a full catalog of spectroscopic sources in this field of view is available in \citetalias{lagattuta2023}). 
The positional parameters ($x,y,\epsilon,\theta$) of the cluster member halos are fixed to the properties of their light distributions. The scaling parameter of the core radius, which has been shown to have a minimal effect on mass models (\citealt{limousin2007,eliasdottir2007}) is fixed to 0.10 kpc according to the parameters of a reference galaxy $L^*$ at the redshift of the cluster, following the mass/light Faber-Jackson scaling relations \citep{faberjackson1976}. The reference magnitude used for $L^*$ in this work is $18.57$ in the z-band. The truncation radius and normalization ($r_\mathrm{cut}$ and $\sigma_0$) pivot parameters are optimized by the model.
The best-fit model is identified as the model with the lowest image plane rms between the observed and model-predicted arc positions. We use a global positional uncertainty of $0\farcs20$ for the observed images in this work. This value was selected to yield a minimum $\chi^2$ relative to the number of degrees of freedom in the final model ($\nu$), such that $\chi^2/ \nu\sim1$. Uncertainties of model parameters and lens model outputs are derived from the MCMC exploration of the parameter space. In this paper, we take great care in evaluating the convergence criteria and statistical robustness of the model, which we describe in~\autoref{sec:convergence}.


\begin{table*}[ht] 
\centering
\begin{tabular}{cllcl}
  \hline
 ID & R.A. [J2000] & Decl. [J2000] & $z$ \\   
  & \arcsec & \arcsec & \\
 \hline
5A 1.1 & 69.2918525 & 0.7297251 & $3.26^{+0.15}_{-0.13}$   \\
5A 1.2 & 69.2922417 & 0.7300867 &   '' \\
5A 2.1 & 69.2918912 & 0.7293545 &   '' \\
5A 2.2 & 69.2923629 & 0.7297084 &   '' \\
5A 3.1 & 69.2918321 & 0.7293301 &   '' \\
5A 3.2 & 69.2924096 & 0.7297634 &   '' \\
5A 4.1 & 69.2921342 & 0.7293573 &   '' \\
5A 4.2 & 69.2922154 & 0.7294140 &   '' \\
5A 5.1 & 69.2921204 & 0.7289892 &   '' \\
5A 5.2 & 69.2922346 & 0.7290551 &   '' \\
12.4 & 69.288877 & 0.729035 & 4.7668 \\
14.1 & 69.290593 & 0.734422 & $5.27^{+0.51}_{-0.20}$  \\
14.2 & 69.283098 & 0.732563 &   '' \\
14.3 & 69.288732 & 0.727831 &   '' \\
14.4 & 69.296070 & 0.729268 &   '' \\
15.1 & 69.290713 & 0.730119 & 3.301  \\
15.2 & 69.290853 & 0.729419 &   '' \\
15.3 & 69.281496 & 0.733488 &   '' \\
151.1 & 69.290714 & 0.729942 & 3.301  \\
151.2 & 69.290789 & 0.729554 &  '' \\
151.3 & 69.281508 & 0.733518 &  ''  \\
16.1 & 69.292577 & 0.731676 & $3.43^{+0.12}_{-0.11}$   \\
16.2 & 69.292990 & 0.730517 &   '' \\
161.1 & 69.292704 & 0.731606 & $3.43^{+0.12}_{-0.11}$   \\
161.2 & 69.293028 & 0.730698 &  '' \\
17.1 & 69.294070 & 0.731823 &  3.53  \\
17.2 & 69.294099 & 0.731139 &  '' \\
17.3 & 69.294411 & 0.730841 &  ''  \\
17.4 & 69.294701 & 0.731381 &  ''  \\
17.5 & 69.292936 & 0.733521 &  ''  \\
17.6 & 69.282188 & 0.733888 &  ''  \\
171.1 & 69.294098 & 0.731862 &  3.53  \\
  \hline
\end{tabular}
\begin{tabular}{cllcl}
  \hline
 ID & R.A. [J2000] & Decl. [J2000] & z \\ 
   & \arcsec & \arcsec & \\
  \hline
171.2 & 69.294133 & 0.731127 &  ''  \\
171.3 & 69.294483 & 0.730862 &  '' \\
171.4 & 69.294726 & 0.731358 &  ''  \\
171.5 & 69.292951 & 0.733557 &  ''  \\
171.6 & 69.282208 & 0.733917 &  ''  \\
18.1 & 69.282352 & 0.730317 & $4.41^{+0.22}_{-0.27}$   \\
18.2 & 69.288898 & 0.725966 &  ''  \\
18.3 & 69.294345 & 0.726562 &  ''  \\
19.1 & 69.283634 & 0.726803 & $2.78^{+0.06}_{-0.07}$   \\
19.2 & 69.287163 & 0.724909 &  ''  \\
19.3 & 69.292788 & 0.724183 &  ''  \\
20.1 & 69.295497 & 0.735368 &  $2.91^{+0.07}_{-0.06}$  \\
20.2 & 69.292331 & 0.737317 &  ''  \\
20.3 & 69.286720 & 0.737837 &  '' \\
c21.1 & 69.294753 & 0.736787 &  \nodata  \\
c21.2 & 69.293960 & 0.737276 &  \nodata  \\
22.1 & 69.292071 & 0.734594 & $2.52^{+0.07}_{-0.07}$  \\
22.2 & 69.283301 & 0.733990 &  ''  \\
22.3 & 69.295179 & 0.731412 &  ''  \\
23.1 & 69.284176 & 0.729843 & $4.99^{+0.28}_{-0.30}$   \\
23.2 & 69.286619 & 0.727473 &  ''  \\
23.3 & 69.296448 & 0.727231 &  ''  \\
24.1 & 69.284980 & 0.729646 & $1.49^{+0.03}_{-0.03}$   \\
24.2 & 69.287365 & 0.727671 &  ''  \\
24.3 & 69.294670 & 0.727653 &  ''  \\
25.1 & 69.291712 & 0.734046 & $3.28^{+0.10}_{-0.10}$  \\
25.2 & 69.294931 & 0.730091 &  ''  \\
25.3 & 69.289714 & 0.728444 &  ''  \\
25.4 & 69.282623 & 0.733195 &  ''  \\
26.1 & 69.283732 & 0.728582 & $4.49^{+0.18}_{-0.14}$   \\
26.2 & 69.286609 & 0.726345 &  ''  \\
26.3 & 69.295396 & 0.725996 &  ''  \\
   \hline
\end{tabular}
\caption{Coordinates and redshifts for systems $14 - 26$, which are the 13 new lensed galaxies that we identify with JWST, as well as the coordinates for image 12.4 and Source \#5A, which we include as new modifications to the existing lensing constraints. The original 13 systems of galaxies are described in \citetalias{lagattuta2023}, and the coordinates we use for them in this model can be accessed from a machine-readable table that accompanies this manuscript. We identify one additional candidate system not used in the lens model with the letter `c'. Redshifts without error bars are spectroscopic redshifts from MUSE, while redshifts with error bars are estimated from the model. All arcs used in the model are available in a separate supplementary file attached to this paper.} \label{tab.arcs}
\end{table*}

\subsection{Multiple Images}
Using a combination of HST imaging and VLT/MUSE spectroscopy, \citetalias{lagattuta2023} presented 13 spectroscopically-confirmed lensed sources (including 9 individual clumps in two of the galaxies) comprising a total of 44 multiple images.

In this work, we detect 13 new multiply-imaged galaxy systems with JWST imaging, doubling the number of modeling constraint families. In addition, thanks to the improved resolution of JWST we identify 57 unique clumps (largely stellar emission knots) in 8 different galaxies. Each clump is well-detected in the JWST pass-bands, as shown in \autoref{fig.clustermodel}. Combining the original and new lens systems (and including individual clumps), our updated model contains a total of \nconstraints\ arcs.

While all \citetalias{lagattuta2023} systems have a confirmed VLT/MUSE redshift, many of the new systems are too faint to be detected with archival spectroscopy. Instead, we include the redshifts of these systems as free parameters in the model. Additionally, we slightly adjust the positions of 12 of the original arc families by aligning them onto clear  
continuum sources that are visible in JWST but undetected in HST. Source \#5, which has a spectroscopic redshift of $z=3.5296$, has an extended tail of emission visible in JWST that we tentatively classify as a separate lensed system: Source \#5A (see~\autoref{sec:sys5}). We perform a slight realignment of Source \#10, which is not visible in JWST (see~\autoref{sec:source10}), to the VLT/MUSE emission centroids. We also robustly identify the positions of the fifth images of Source \#1 and Source \#2, as well as the third image of Source \#5, which were unclear in \citetalias{lagattuta2023} since these less-magnified images were not clearly visible in HST, likely due to magnification effects. 

We highlight the presence of Source \#17, a lensed galaxy affected by both the cluster potential and localized galaxy-galaxy lensing effects from cluster member perturbers. This lensing configuration results in six images of the galaxy, where two distinct sub-clumps can be identified in each image. The overall lensing configuration of the source is likely a naked cusp \citep{lewis2002}, but the cluster member located near images 17.1-17.4 splits one image of this source into four. We separately parameterize the properties of this cluster member in the model to account for this contribution. Further analysis of this source will be presented in L. Fung et al. (in prep).

We revisit the existing VLT/MUSE data to search for spectral features in the 13 new lensed sources identified in this work. The spectra of all the multiple images that are within the MUSE field of view are extracted and co-added together to search for additional emission or absorption features. The co-added spectra of Source \#15 show a significant CIII] doublet, giving a source redshift of $z=3.301$. Source \#17 shows a weak damped Lyman-$\alpha$ feature, as well as several weak silicon absorption features that seem to be separate from any contamination from sky noise or nearby cluster members. These features appear in three out of the six images of this galaxy, and imply a source redshift of $z=3.523$. We implement both of these spectroscopic redshifts into the model. 

We also make one new spectroscopic identification of a fourth counter-image for Source \#12. Three images were originally reported in \citetalias{lagattuta2023}; the JWST imaging suggests the presence of a fourth image near the predicted location from the \citetalias{lagattuta2023} model. The spectrum at the location of this counter-image shows a double-peaked Lyman-$\alpha$ feature of the same intensity as the other three images of Source \#12.

We show the spectra for each of the new detections made in this work in~\autoref{sec:newspecz}. No other spectroscopic features could be identified from the VLT/MUSE data for the rest of the new sources, and so we optimize the redshifts of the remaining 10 sources within the model. We provide the coordinates and model estimates for the redshifts or the new lensed galaxies identified in this work in~\autoref{tab.arcs}. The full list of coordinates for all arcs used in the model is provided in an additional machine-readable table included in the online materials for this work.

\subsection{Model Optimization} \label{sec:modelopt}
In this section, we introduce the iterative process of modeling this cluster, and describe the various test models that were developed to enable efficient optimization of the many arcs identified from the JWST imaging. We dub the final model the `fiducial model', and present it in in~\autoref{sec:convergence}.

The large number of constraints available to the lens model is at once beneficial, in that the presence of a lensed arc within a region is a direct constraint on the local lensing potential in that region; and computationally difficult, as achieving convergence on a \lenstool~model with this many constraints requires an approach that both minimizes the rms and allows enough flexibility for the model to avoid optimizing around local minima, rather than the parameters that describe the `true' best model. This computational complexity initially prevented us from achieving convergence when running a lens model using the complete set of lensing constraints with broad, uniform priors on the optimized model redshifts and halo parameters. While the initial best-fit model, which is parameterized in the same way as the fiducial model, provided a reasonable solution, the parameter space was highly undersampled and could not be used to derive meaningful uncertainties. This insufficient sampling was tied to the use of broad priors, which caused the sampler to spend an insufficient amount of computational time exploring the space around the best-fit model, such that in some cases the best-fit values fell well outside of the range that was more extensively well-sampled by the MCMC walkers. We limited the amount of computational time given to the sampler following the work presented in \cite{limousin2025}, with a maximum allowed runtime of three weeks on a modern 12-24 core machine. 
The solution to create a well-sampled and well-converged model was two-fold. First, we generated abridged models (see below) using fewer arcs to identify appropriate priors for each parameter, which reduced the amount of computational time needed to search the parameter space for the global minimum; and second, we experimented with varying two parameters of the \lenstool~MCMC sampler: the \texttt{RATE} parameter, which dictates the speed of convergence in the MCMC sampler \citep{jullo07}; and the \texttt{Nb} parameter, which corresponds to the number of iterations in the MCMC chain after the burn-in stage. A full description of these parameters is provided in Section 4.1 of \cite{limousin2025}. 

To efficiently use computational time, we created two abridged models: one using only lensed galaxies with spectroscopic redshifts (Model~S); and one using all the lensed galaxies identified in this paper, but with a maximum of only two clumps included for each multiple image (Model~C). Model~S was designed to remove the time-consuming optimization of the redshifts for lensed galaxies with no spec-$z$, and contained a total of 232 arcs. Model~C reduced the computational complexity in the inner region of the cluster, and contained a total of 101 arcs. We then examined the distribution in the parameter space of each parameter in Model~C and Model~S. In each case, the parameter space exploration of both abridged models overlapped with a high degree of significance. All input parameters and model components were left equal in both Model~C and Model~S. 

We selected boundaries for the priors based on the overlapping distributions of Model S and Model C, where we took the lowest bound from either model as our lower limit on the prior, and the highest bound from either model as our upper limit on the prior. 
We then ran our final model, which included all \nconstraints~arcs presented in this work, using a uniform prior that we set to these narrower limits. Narrowing the allowable parameter space of the model in this way allowed the model to achieve convergence in 7 days on the Great Lakes HPC cluster using a total of 16 CPU cores, which is an improvement on the original computational time of the model (in excess of 20 days). 

Once we had a converging model, we experimented with lowering the \texttt{RATE} and increasing the Nb parameter of the \lenstool~MCMC sampler to see if this would affect results. We ran models with (\texttt{RATE},\texttt{Nb}) set to (0.05,500), (0.05,1000), (0.05,1500); and (0.1,500), (0.1,1000), and (0.1,1500). We find that the parameters in each model are stable regardless of the changes to \texttt{RATE}, but low values of \texttt{Nb} would sometimes insufficiently sample the parameter space around the best-fit model. In our final models, we set the \texttt{RATE} to 0.1, and \texttt{Nb} to 1500 to achieve good sampling.

\subsection{Fiducial Model}\label{sec:convergence}

A major goal behind remodeling this cluster was to explore how the new JWST arcs affect our optimization of the mass distribution of the cluster. The original \citetalias{lagattuta2023} model, which was also constructed with \lenstool, required an external shear component and a light-unaffiliated mass clump  \citep[LUMC;][]{limousin2025+lumc}, also known as a `dark clump', to effectively reproduce the observed constraints. In the new JWST model, we find that these components, which are difficult to interpret physically since they are unassociated with optical counterparts, are not strictly necessary to produce a model capable of reproducing the constraints. We tested models with and without these components, and we discuss the effects of incorporating them into the lens model in~\autoref{sec:extshear}. However, we do not include them in the main result we present in this paper -- the fiducial model -- since they do not result in a significant improvement in the rms. The fiducial model has \nconstraints~arcs and 45 free parameters, and results in an image plane rms of \finalrms. The best-fit parameters are tabulated in \autoref{tab.bestfitparams}.

We evaluate models using the Bayesian Information Criterion (BIC; \citealt{schwarz1978}) and Akaike Information Criterion (AIC; \citealt{akaike}) metrics. These criteria penalize additional model complexity (i.e., greater number of free parameters) that yields no notable improvement in the goodness of fit. The model with the lowest BIC and AIC value is statistically preferred \citep{jauzacharveymassey,hareyjauzacmassey}.

The fiducial model is constructed using seven parametric halos that are centered on observable stellar counterparts, as well as halos centered on cluster members. The parametric halos are optimized as follows: one cluster-scale DM halo, two group-scale DM halos, two galaxy-scale halos centered on the BCG, and two galaxy-scale halos to optimize cluster members which have a significant local perturbation effect on nearby lensed galaxies. The two group-scale DM halos are optimized to be smaller than the cluster-scale halo, but larger than individual galaxy-scale halos, and are placed on the second and third brightest galaxies in the field to account for additional mass in these regions. We report the best-fit parameters of the fiducial model in~\autoref{tab.bestfitparams}.

The core radius $r_{\mathrm{core}}$ of the main cluster DM halo is quite large, implying a cored mass distribution. We tested this result by running a test model where we restricted the optimization of the core radius to be less than 10 kpc, which forced the model to try to reproduce the observational constraints using a non-cored DM halo \citep{limousin2022}. This non-core mass model resulted in an rms of $2\farcs84$, which is an increase of $2\farcs61$ over the fiducial model. The large difference in the rms strongly indicates that a cored mass model is favored for this cluster. 

Adding the two additional group-scale halos in the North-west and South-east allows the model to successfully reproduce the fifth image of the eastern H-U system (Source \#2), as well as the third image of Source \#5. Removing these two halos from the model results in an rms of $0\farcs38$ and a statistically worse optimization: the BIC and AIC values of the model were over ten times higher than the fiducial model when they were removed than when they were included. We thus keep these components in the parameterization of the model. 

Of the two halos used for the BCG, one is fixed to the measured centroid, ellipticity, and position angle of the stellar mass as measured from the galaxy's light, allowing the slope ($r_\mathrm{core}$, $r_\mathrm{cut}$) and normalization ($\sigma_0$) to vary. The other is fixed to the position of the BCG and is allowed to vary in both shape and size. This extra halo accounts for any additional mass within the center of the cluster that may originate from, e.g., the DM of the cluster that is not captured by the cluster-scale DM halo, and/or the central DM distribution associated with the BCG. The scale of the second BCG halo is quite large, but its position angle and ellipticity are different than the cluster-scale DM halo. This gives the central mass distribution of the cluster a boxier shape than a single dPIE halo can produce, similar to the parameterization used in other clusters with complex mass distributions  \citep[e.g., Abell 370;][]{lagattuta2019}. 
Optimizing only one halo around the BCG results in a worse statistical performance of the model, with BIC and AIC values 1.6 times higher than the fiducial model. 

Finally, the two galaxy-scale halos are included to account for the localized lensing effects on the background source galaxies \#13 and \#17. 

With this parameterization of the cluster mass, we are able to achieve an average rms of $0.233\pm0.006$, where the best-fit model has an rms of \finalrms. We measure the standard deviation of the rms by considering all runs in the MCMC chain. By contrast, \citetalias{lagattuta2023} reports an average rms of $0.309\pm0.008$, where the best-fit model has an rms of $0\farcs29$. The improvement in the rms is influenced by the number of new constraints, which significantly tighten the allowable parameter space explored by the model. Crucially, the new model achieves a closer alignment between the best-fit rms and the average rms without using either an external shear component or a LUMC, which were both elements of the \citetalias{lagattuta2023} model. We explore the addition of a shear component to the fiducial model, but only find a small additional improvement in the rms (a difference of $0\farcs03$) and the BIC and AIC (1.8 times lower than the fiducial model). We do not robustly identify any physical components that could explain the presence of this shear term, and since the improvements are marginal and may be associated with the limitations inherent to parametric lens modeling, we do not include a shear term in the fiducial model. We discuss the impact of using a residual shear term and the \citetalias{lagattuta2023} LUMC more thoroughly in \autoref{sec:extshear}.

\begin{deluxetable}{l|lcccccccc}
\tablecaption{Best Fit Lens Models \label{tab.modelstats}}
\tablehead{ 
    \colhead{Model} &
    \colhead{$\nu$} & 
    \colhead{$\mathrm{N_{Free}}$} & 
    \colhead{$\chi^2$} & 
    \colhead{$\chi^2_{\nu}$} &
    \colhead{$\mathrm{log}(L)$} & 
    \colhead{BIC} & 
    \colhead{AIC} & 
    \colhead{$\mathrm{rms}$} 
}
\startdata
Fiducial & 355 & 45 & 341 & 0.96 & 210 & -149 & -107 & \finalrms \\ 
\hline
Model S & 346 & 28 & 1439 & 4.52 & -77 & 318 & 139 & 0\farcs25 \\
Model C & 142 & 39 & 750 & 7.28 & -95 & 384 & 204 & 0\farcs27 \\
\enddata
\tablecomments{ A comparison of statistical properties for the mass models presented in this paper. $\nu$ corresponds to the degrees of freedom (d.o.f), which is the difference between the number of arcs and the number of free parameters ($N_{\mathrm{Free}}$). The best-fit $\chi^2$ and the reduced $\chi^2$, which represents the $\chi^2$ divided by the d.o.f., are given in the fourth and fifth columns, respectively. The best-fit log likelihood, BIC, AIC, and rms in the image plane are also provided.}
\end{deluxetable}

\begin{figure*}
\begin{minipage}{1\linewidth}
\centering
    \includegraphics[width=0.3\linewidth]{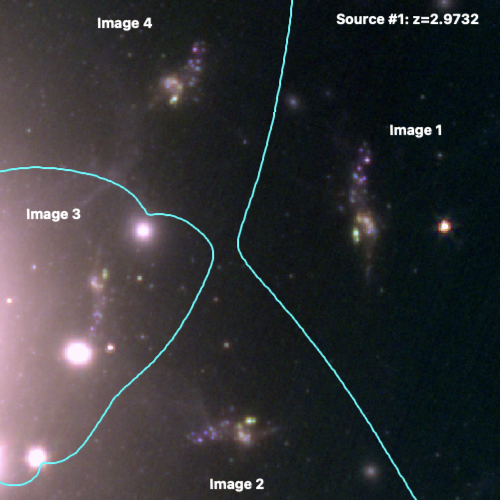}
    \includegraphics[width=0.3\linewidth]{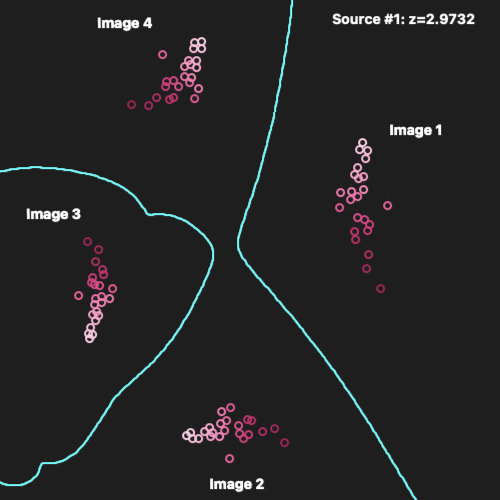}
    \includegraphics[width=0.3\linewidth]{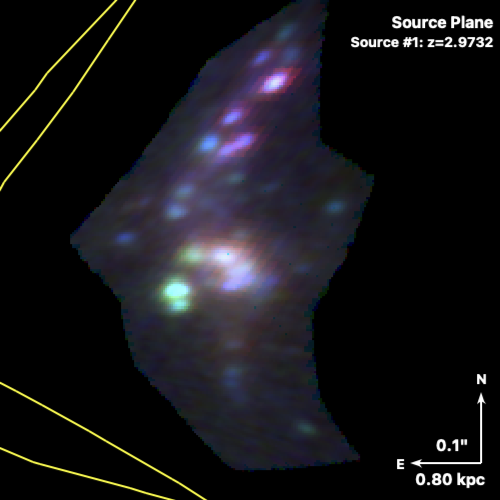}
\end{minipage}\\
\begin{minipage}{1\linewidth}
\centering
    \includegraphics[width=0.3\linewidth]{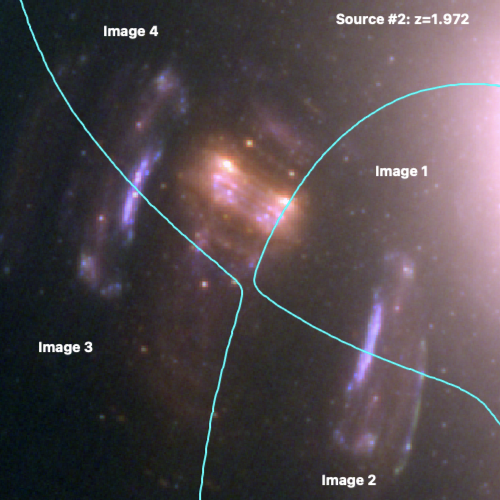}
    \includegraphics[width=0.3\linewidth]{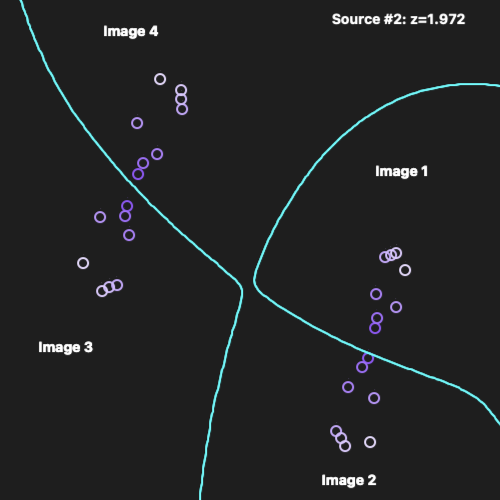}
    \includegraphics[width=0.3\linewidth]{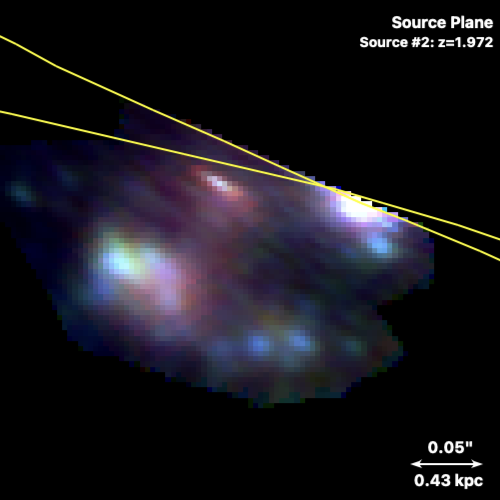}
\end{minipage}
\begin{minipage}{1\linewidth}
\centering
    \includegraphics[width=0.3\linewidth]{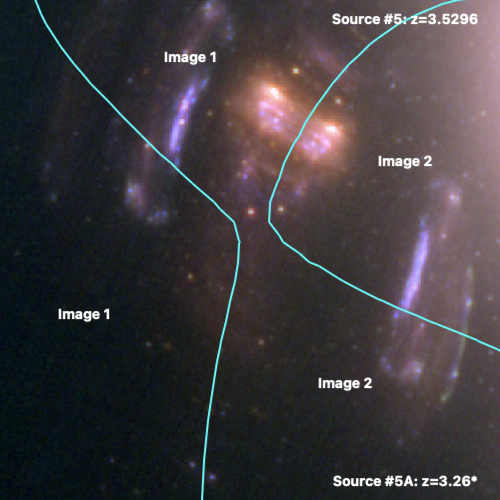}
    \includegraphics[width=0.3\linewidth]{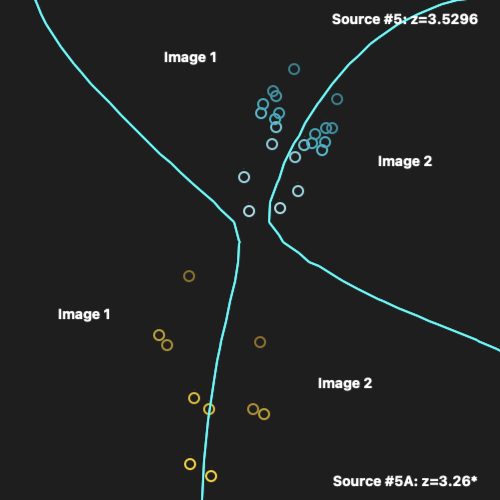}
    \includegraphics[width=0.3\linewidth]{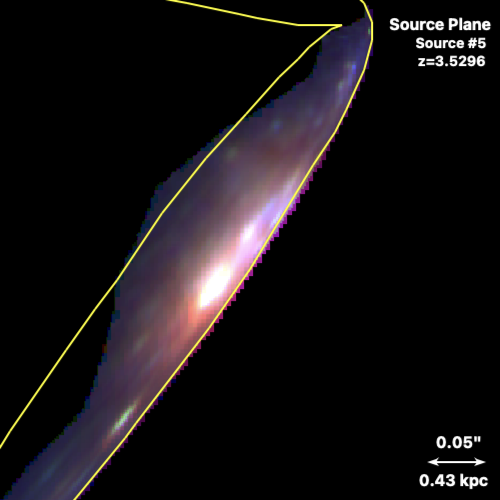}
\end{minipage}
    \caption{
    Zoom-in view of the central images of sources \#1 (\textit{top row}), \#2 (\textit{middle row}) and \#5+5A (\textit{bottom row}). Sources \#2 and \#5 form the `Cosmic Mantis'. Each row shows a JWST color composite image, using the same NIRCam bands as ~\autoref{fig.clustermodel}, in the left panel -- scaled to highlight the features in the system, and a visual guide in the middle panel. The critical curves of the fiducial model are plotted in cyan for a source redshift as labeled. The colored circles indicate the positions of the clumps that were used as constraints, color-coded to show the mapping from image to image and to guide the eye to the lensing configuration and symmetry of the system. 
    The critical curve in the bottom panels is plotted for the spectroscopic redshift of Source \#5 - $z_{\mathrm{spec},5}=3.5296$. The model redshift for Source \#5A is $z_{\mathrm{model},5A}=3.26$, but we do not show this critical curve in the plot for visual clarity. In the right-most panel, we show the source plane reconstructions for Source 1, using image 1; Source 2, using image 1; and Source 5, using image 2. The source plane caustics, which map to the critical curves in the image plane, are shown as yellow lines.
    }
    \label{fig:HUsnapshots}
\end{figure*}

\section{Results} \label{sec:results}

 \subsection{Model Outcomes}

The optimization of the best-fit fiducial model yields an rms of \finalrms~using the \nconstraints~arcs. When considering the rms of only the visible H-U systems (Sources \#1 and \#2) and Source \#5, which are the most highly magnified galaxies in the field and which have the highest amount of visible substructure and lensing constraints in each multiple image, we measure an rms of $0\farcs13$ across 193 arcs. We thus achieve a highly accurate reproduction of the observed multiple images in the clumpiest galaxies in this field.

Most of the mass within the model is concentrated within the cluster DM halo and the second halo placed on the BCG. The velocity dispersions of these halos are moderately degenerate with each other, while $\epsilon$ and $\theta$ are weakly correlated; no other parameters show a strong correlation. Removing this halo and increasing the mass of the cluster DM halo worsens the statistical performance of the model (see discussion in~\autoref{sec:convergence}) and causes the rms of the H-U systems and Source \#5 to worsen by $0\farcs1$, which we consider sufficient evidence to justify the inclusion of this halo. 

We compare the basic properties of our fiducial model to the model presented in \citetalias{lagattuta2023}. Similarly to \citetalias{lagattuta2023}, the largest mass components in the model are the cluster-scale DM halo and the BCG halo, with the addition of the new DM halos centered on the second and third brightest cluster members, which are located in the North-west and South-east of the cluster, respectively.

The inclusion of the DM halo in the North-west allows the less magnified multiple images in this region of the cluster to be reproduced by the model. Removing this halo causes the rms to increase by $0\farcs1$ and significantly increases statistical uncertainties. The DM halo in the South-east further improves the model's rms by about $30\%$, and also eliminates the need for the LUMC used in the \citetalias{lagattuta2023} model, which was also located in the southern part of the cluster (see~\autoref{sec:extshear} for further discussion). 

The ellipticity of the cluster-scale and BCG halos is moderate ($\epsilon\sim0.3-0.55$), while their position angles and spatial centroids are aligned, in line with the results from \citetalias{lagattuta2023}. 

The mass profile also does not change significantly from \citetalias{lagattuta2023}. In the fiducial model, we measure the total integrated mass of the cluster within 20 kpc to be $5.09\pm0.14\times10^{12}M\odot$; the mass within 110 kpc to be $8.22\pm0.44\times10^{13}M\odot$; and the mass within 900 kpc to be $5.26\pm0.04\times10^{14}M\odot$, which represent a $<1\%$ increase, $16\%$ decrease, and $47\%$ increase from the mass reported in \citetalias{lagattuta2023}, respectively. 

The inner mass measurement is the most consistent with \citetalias{lagattuta2023}, but the extended set of constraints we employ requires more mass to be present in the outer regions of the cluster around the second and third brightest cluster member galaxies, which accounts for the difference in the measurements.

\begin{deluxetable*}{l|lcccccccc}
\tablecaption{Best Fit Parameters\label{tab.bestfitparams}}
\tablehead{ 
    \colhead{Model} &
    \colhead{Component} & 
    \colhead{$\Delta x$ ($''$)} & 
    \colhead{$\Delta y$ ($''$)} & 
    \colhead{$\epsilon$} &
    \colhead{$\theta$ ($^\circ$)} & 
    \colhead{$r_{\mathrm{core}}$ ($''$)} & 
    \colhead{$r_{\mathrm{cut}}$  ($''$)} & 
    \colhead{$\sigma$ (km $s^{-1}$)}
}
\startdata
Fiducial & Cluster DM Halo  & $-0.41_{-0.01}^{+0.06}$ & $1.71_{-0.21}^{+0.19}$ & $0.31_{-0.02}^{+0.01}$ & $86.8_{-3.2}^{1.9}$ & $12.4_{-0.5}^{+0.6}$ &  [349]   & $657_{-25}^{+17}$ \\
$N_{\mathrm{Arcs}}$: \nconstraints& BCG Halo 1     &  [0.0]   & [0.0]    &  [0.29]   & [101] & $0.08_{-0.1}^{+0.2}$ & $15_{-3}^{+4}$ & $260_{-3}^{+7}$ \\
 & BCG Halo 2 & [0.0]    & [0.0]    & $0.73_{-0.03}^{+0.01}$ & $112.3_{-0.8}^{+0.3}$ & $12.6_{-0.6}^{+0.1}$ & $345_{-32}^{+1}$ & $666_{-18}^{+24}$ \\
 &  North-west Halo & [24.54]    & [30.62]    &  $0.08_{-0.03}^{+0.01}$   &   $120.6_{-3.4}^{+3.7}$ & $23.3_{-2.3}^{+1.8}$ & $210_{-29}^{+100}$ & $313_{-19}^{+18}$ \\
&  South-east Halo  & [12.65]    & [-31.67]    &  $0.6_{-0.1}^{+0.1}$   & $141.0_{-10.5}^{+9.5}$ &   $10.0_{-0.1}^{+3}$ & $201_{-73}^{109}$ & $102_{-9}^{+16}$ \\
&   Cluster Member 1  &  [-17.04]   &  [0.57]   & [60.0]    &  [0.25]   & $0.01_{-0.01}^{+0.02}$ & $2_{-0}^{+1}$ & $128_{-5}^{+5}$ \\
&   Cluster Member 2   &  [18.49]   &  [-14.81]   & $0.54_{-0.13}^{+0.01}$ & $116.4_{-17.1}^{+16.3}$ & $0.3_{-0.2}^{+0.1}$ & $1_{-0}^{+0}$ & $203_{-11}^{+25}$ \\
&    \lstar & \nodata & \nodata & \nodata & \nodata & \nodata & $23_{-2}^{+1}$ & $108_{-2}^{+3}$ \\
\hline 
\enddata
\tablecomments{ Optimized parameters for the best-fit fiducial model. $\Delta x$. and $\Delta y$ are defined in relation to the BCG of the clusters. Position angles are measured north of west, and the ellipticity $\epsilon$ is defined as $(a^2-b^2)/(a^2+b^2)$. Error bars correspond to the $68\%$ confidence level. Non-optimized values are indicated by brackets. The properties of the reference galaxy used to scale the cluster member mass parameters are given in the row designated \lstar. 
}
\end{deluxetable*}

\subsection{Source Reconstruction}

The proximity of Sources \#1, \#2, and \#5 to the critical curve of the lensing model indicates that they are very strongly magnified {(by $\sim25$x, $\sim100$x, and $\sim150$x, respectively)}, and in the case of Sources \#1 and \#2, are essentially undistorted, highly-magnified images of their source galaxies. To test this unique property of H-U configurations, we reconstruct the original morphology of the lensed galaxy at $z=2.9732$ for Source \#1, $z=1.9722$ for Source \#2, and $z=3.5296$ for Source \#5, by ray-tracing individual pixels associated with each galaxy in the image plane back to the source plane. This transformation is applied using the lens equation, $\vec{\beta} = \vec{\theta} - \vec{\alpha}(\vec{\theta})$, where $\vec{\beta}$ is the source plane position of the pixel and $\vec{\theta}$ is the observed position. $\vec{\alpha}$ is the deflection that occurs due to the presence of the lens, and it is scaled by $d_{LS}/d_S$, where $d_{LS}$ is the angular diameter distance between the lens and the source and $d_S$ is the angular diameter distance between the observer and the source. 

The rightmost panels of \autoref{fig:HUsnapshots} show the reconstructed source images for each of these galaxies, as well as the magnification for each source at their corresponding redshifts in the image plane. The source reconstruction of Source \#1 appears very similar to its lensed image, indicating that individual components of the galaxy are being magnified isotropically. Source \#2 is more compressed due to the merger of two images happening along the critical curve, which causes only part of the source galaxy to be lensed; however, the piece that is lensed is very similar between the source and image planes. The H-U images thus show an undistorted view of the morphology of the galaxy that is not strongly sensitive to systematics in the lens model, and thus its morphological interpretation is straightforward and more robust compared to highly distorted giant arcs. 

Source \#1 is of particular interest in this regard due to the strong levels of color variation present between its individual clumps. Each image of this galaxy shows both green clumps, or `hot knots', which are highly ionized [OIII] emission regions that may also be potential lines of sight for Lyman-continuum escape \citep{rivera2026}; darker red regions, which are either old, dusty, or both; and an orange `proto bulge' forming in the bottom right. This indicates that multiple stellar populations are simultaneously resolved within each image of the galaxy. The same type of color variation can be seen in the red and blue clumps of Source \#2. Similarly, although the reconstructed image of Source \#5 is more distorted relative to its appearance in the image plane since it is not an H-U system, definitive color variation is clearly visible in its substructure clumps, with the orange bulge at the center, the green knot to the north, and the green, red, and blue knots to the south.

\subsection{Clumps in Multiple Images}
Luminous substructure clumps are a particularly useful feature in lens modeling, as they subdivide each multiple image into discrete pieces that can be easily mapped between each image, which places tighter spatial limitations on the model's reproduction of the observed multiple images \citep{limousin2026}. The high resolution of the NIRCam imaging enables the identification of small substructure clumps in each of the visible H-U systems, Sources \#1 and \#2, as well as in Source \#5. \autoref{fig:HUsnapshots} shows each of these three systems in more detail. We select the most reliable clumps in each multiple image, which we define as the clumps that appear in each of the most highly-magnified images, to use as constraints in the lens model, with the requirement that each clump must appear in each of the highly magnified images near the center of the cluster. In total, we identify 24 clumps in Source \#1, 8 in Source \#2, and 16 in Source \#5. Of the Source \#5 clumps, 11 lie in the northern part of the galaxy identified in L23, along with 5 located in a new, extended southern region only clearly visible in the JWST data (see the bottom row of \autoref{fig:HUsnapshots}). The radial and tangential critical curves describe the regions in the image plane along which images experience near-infinite magnification in either the tangential or radial directions. At the redshifts of all three sources, these curves are nearly touching, which is expected from the H-U systems (\citealt{petters2001}), but is perhaps more curious in Source \#5, which does not have the morphology of an H-U galaxy since it only produces two multiple images in this region instead of four. Instead, the top and bottom portions of Source \#5 rest along the radial and tangential critical curves, respectively, which strongly boosts their magnification and makes them much brighter compared to the other non H-U sources in the field. We discuss the possibility that Source \#5 may in fact be two separate lensed galaxies at slightly different redshifts in \autoref{sec:sys5}.

We measure the magnifications of the clumps of the two visible H-U systems (Sources \#1 and \#2), as well as for Source \#5, from the best-fit lens model. We calculate the errors for these measurements from a random selection of 100 models drawn from the posterior distribution. In Source \#1, we find magnifications of between $\mu\sim8-33$ for the central four images. In Source \#2, we find magnifications of between $\mu\sim27-210$ for the central four images. In Source \#5, we find magnifications of between $\mu\sim80-440$ for the central two images. We provide a description of the errors for these measurements and a truncated list of all magnifications for each clump in~\autoref{sec:clumpmag}; the full list can be accessed from a machine-readable table that accompanies this manuscript.  

We evaluate the impact of the inclusion of these clumps on the lens model by considering a variation of the fiducial model: the `NoClumps' model, which only uses a single constraint for each image centered at the barycenter of the observed light (using one less clump per system than Model C; see ~\autoref{sec:modelopt}) for a total of 84 arcs, but is otherwise identical to the fiducial model. In contrast, the `Clumps' fiducial model uses all distinct morphological points for each system as constraints. As a quantitative assessment, we wish to compare the performance of the fiducial model, which uses all possible clumps (Clumps), to the model without clumps (NoClumps). 

However, since these two models use a different number of arcs, comparing the model-reported rms or $\chi^2$ is not a fair evaluation. Using simulated data, \cite{johnsonsharon2016} showed that while models with a lower number of multiple image systems are less accurate, their rms is artificially low, especially compared to models with significantly more constraints. To avoid this bias, we make a final `prediction' of the image plane rms for both models using both sets of arcs, similar to the approach of \cite{remolina2018}. 

We compare the rms of the Clumps model to the rms of the NoClumps model for the full set of \nconstraints~arcs. We find that the Clumps model achieves an rms of \finalrms, while the NoClumps model has an rms of $0\farcs32$. The rms improves by $32\%$ in the Clumps model. When comparing the performance of these models against the NoClumps set of arcs, we find that the NoClumps model performs slightly better ($7\%$) than the Clumps model for arcs that do not have clumps. 

From these tests, we can see that the model does not dramatically change with the addition of all possible clumps as constraints. The addition of these substructures into the lens model does lead to better accuracy and precision, especially in those systems that have many clumps; notably, we find a $70\%$ improvement in the rms of the H-U systems. This is particularly useful when searching for e.g., small dark matter substructures that may be influencing these clumps (see discussion in~\autoref{sec:dmsubstructure}).

\subsection{Time Delay Predictions}
The amount of likely star-forming substructure and high magnification of the lensed galaxies lends itself to possible transient searches. Detections of transient objects in lensed systems of galaxies can be used to constrain cosmology \citep{napier2023}, provide extra checks on the amount and nature of DM present in the region \citep{diego2018,croon2026}, and evaluate stellar evolution \citep{kelly2015,li2025}. Sources \#1, \#2, and \#5 are the most magnified, morphologically detailed systems in this cluster. We use the lens model to calculate time delays for the appearance of transient objects in each of these systems, which can immediately be used to facilitate future follow-up observations. Predicted time delays for the fiducial model are provided in~\autoref{sec.tdelay}.

\section{Discussion} \label{sec:discussion}

\subsection{Arcs of Interest}
In this subsection, we discuss interesting properties of two source galaxies in this field: Source \#10 and Source \#5. 

\subsubsection{Source \#10}\label{sec:source10}
\begin{figure}
\begin{minipage}{1\linewidth}
 \centering
    \includegraphics[width=0.95\linewidth]{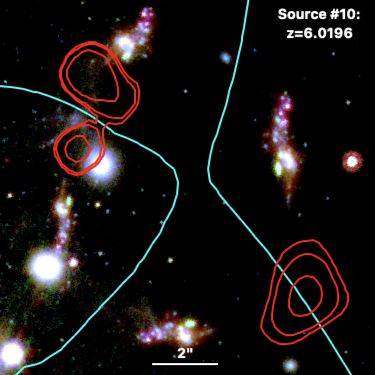}
\end{minipage}
    \caption{\textit{Left: }False color image of a $6\arcsec \times 6\arcsec$ region around Source \#10, constructed using F090W, F200W, and F444W imaging, from which the light from the BCG has been removed using a basic Sersic profile to enhance the visibility of lensed sources in this region. The VLT/MUSE contours for the Lyman-$\alpha$ emission feature of Source \#10 are plotted in red, while the critical curve for the fiducial model at the redshift of the source ($z=6.0196$) is shown in cyan.
    }\label{fig.source10}
\end{figure}

\citetalias{lagattuta2023} confirmed a total of 13 systems of lensed galaxies with spectroscopic redshifts from VLT/MUSE. In this work, we are able to match these identifications to faint features in the JWST imaging for all systems save for one: Source \#10, which is, along with Sources \#1 and \#2, an H-U system. The redshift of this source is $z=6.0196$, which is a spectroscopic measurement made from Lyman-$\alpha$ emission. The source is unique because it displays no visible continuum in any of the NIRCam imaging, as seen in \autoref{fig.source10}. The NIRCam filters at this redshift cover an approximate wavelength range of 1136-7110 \AA. At the depth of our data, the $5\sigma$ limiting magnitude for a point source within a circular aperture of $0\farcs1$ in the four NIRCam bands is 29.5-30.1 magnitudes; any objects fainter than this cannot be detected in imaging with similar properties as the one used in this analysis. The non-detection of this source in rest-frame optical wavelengths is significant given that its proximity to the critical curve gives it a high magnification (on average for the source, $\mu > 40$; this number increases significantly to $\mu>100$ for the two images in the south, which appear to be merging along the critical line).

This result implies that Source \#10 is likely an extremely faint, low-mass, star-forming Lyman-$\alpha$ emitter (LAE). This type of LAE is rare, but not unusual. Recent work by \cite{goovaerts2024} used 15 bands of HST and JWST imaging from the UNCOVER treasury survey \citep{bezanson2024}, which had exposure times between 4-6 hr per filter with an approximate $5\sigma$ limiting magnitude range of 29.5-30.0, to study a total of 154 LAEs detected with MUSE in the galaxy cluster Abell 2744. Of the 154 total MUSE LAEs, only 121 were detected in the UNCOVER photometry. The 33 LAEs that were not detected constitute a non-trivial portion of the sample. Source \#10 may fall into the same class of objects as these non-detected LAEs, but additional exploration of its spectroscopic properties using, e.g., space-based spectroscopy covering a wider range of rest-frame UV/optical wavelengths is needed before any definitive conclusions can be made.

\begin{figure}
\begin{minipage}{1\linewidth}
 \centering
    \includegraphics[width=1\linewidth]{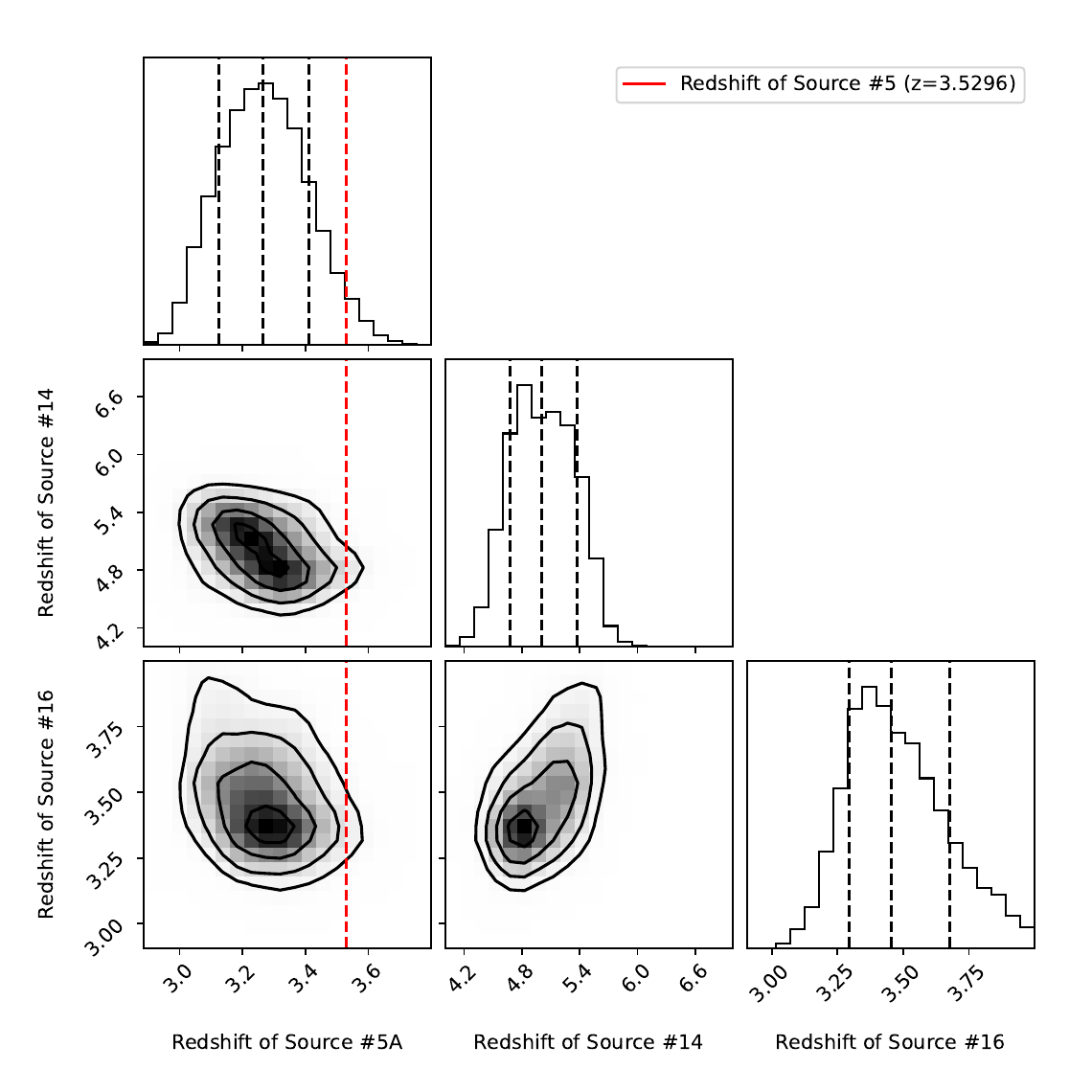}
\end{minipage}
    \caption{The optimization of the redshift for the faint portion of Source \#5, labeled here as \#5A, as well as the two model-predicted highest-$z$ sources in the cluster, yields a well-constrained distribution, which is shown in this corner-plot. The spectroscopic redshift of Source \#5 ($z_{\mathrm{spec},5}=3.5296$), which is shown as the dashed red line, falls within the tail end of the distribution for the redshift of Source \#5A, but since the model prediction for Source \#5A favors a lower redshift ($z_{\mathrm{model},5A}=3.26\pm0.14$), Source \#5A may be separate from Source \#5.
    }\label{fig.cornerspaceinvader}
\end{figure}

\begin{figure*}
\centering
\begin{minipage}{0.25\linewidth}
    \includegraphics[width=1\linewidth]{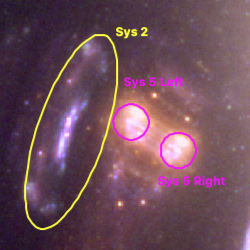}
\end{minipage}
\begin{minipage}{0.33\linewidth}
    \includegraphics[width=1\linewidth]{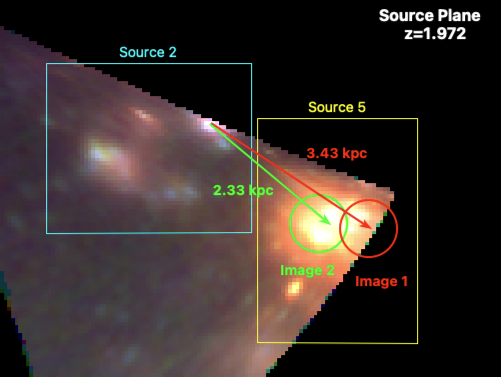}
\end{minipage}
\begin{minipage}{1\linewidth}
    \includegraphics[width=1\linewidth]{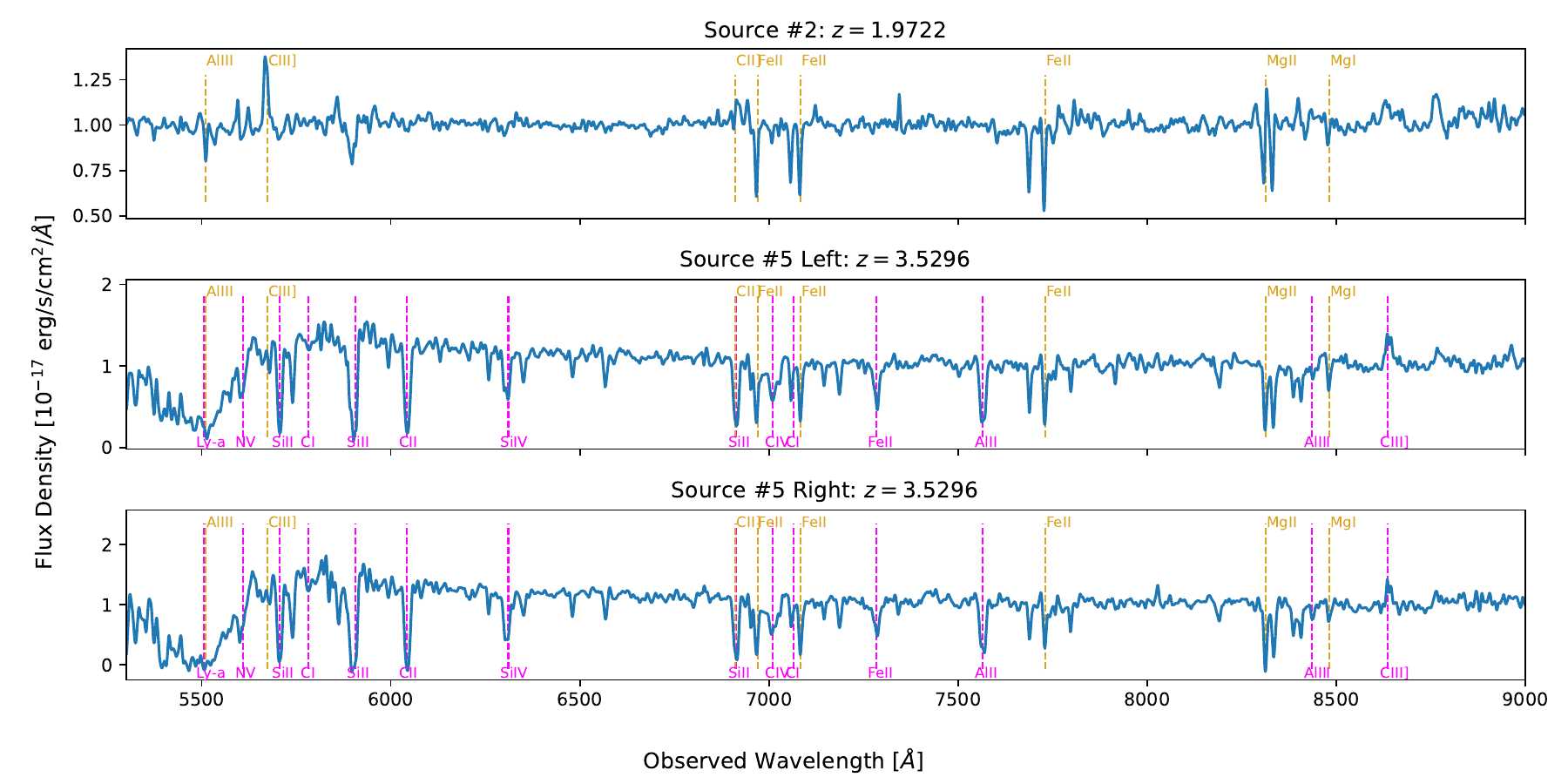}
\end{minipage}
    \caption{
    \textit{Top Left:} False-color JWST image showing H-U Source \#2 and Source \#5. The spectral extraction regions for Sources \#2 and \#5 are shown in yellow and magenta, respectively.
    \textit{Top Right:} We reconstruct the $z=1.972$ plane from the lens model to show the spatial separation between Sources \#2 and \#5 in this plane.
    \textit{Bottom:} Normalized VLT/MUSE spectra for Source \#2 (top) and Source \#5 (middle and bottom), extracted from the regions shown in the right panel. Notable absorption and emission lines are marked in each spectrum with vertical dashed lines; iron and magnesium lines at the redshift of the foreground Source \#2 appear in absorption in the spectrum of Source \#5, and are marked in yellow in both spectra. The absorption lines intrinsic to Source \#5 are labeled in magenta.
    }
    \label{fig:speccompare}
\end{figure*}

\subsubsection{A Single Galaxy, or Two?} \label{sec:sys5}
We draw special attention to Sources \#5 and \#5A (shown in the bottom row of \autoref{fig:HUsnapshots}). Although these sources are spatially close to each other, their features, colors, and surface brightnesses are quite different, indicating that they may be two separate galaxies. The spectroscopic redshift of Source \#5 is $z_{\mathrm{spec},5}=3.5296$. Due to its low surface brightness, we do not have an independent MUSE redshift solution for Source \#5A. Moreover, in our preliminary models the lensing symmetry of the clumps in Source \#5A was moderately inconsistent with the critical curve for the source redshift of \#5. We therefore investigated whether or not \#5 and \#5A are actually two separate galaxies by freeing the optimization of Source \#5A's redshift within the model. The resulting optimized redshift distribution is shown in \autoref{fig.cornerspaceinvader}. The predicted redshift for Source \#5A favors a lower redshift in the fiducial model, which gives a redshift of $z_{\mathrm{model},5A}=3.26\pm0.14$. The statistical error of this measurement barely includes the spectroscopic redshift of Source \#5.  If these are indeed two galaxies at different source planes, then this region of the cluster is of particular interest because it contains 5 separate galaxies (Sources \#5, \#5A, \#2, \#15, and \#16), which cover a redshift range of $1.97-3.53$ and generate a total of 12 multiple images within $\sim50$ kpc in projection from the cluster core. Even if Source \#5 and Source \#5A are parts of the same galaxy, the presence of this many lensing observables places extremely fine constraints on the mass within this region, enabling precision measurements of, e.g. DM substructure. However, until spectroscopic confirmation of the redshift for this source is achieved, it will likely be better to remove this source when considering DM substructure at the $10^9 M_\odot$ level, as its inclusion may affect the positional uncertainties measured for the clumps of its central two images (see discussion in ~\autoref{sec:dmsubstructure}).

\subsection{Arc Tomography}

The spectroscopy and photometry of Sources \#2 ($z=1.972$) and \#5 ($z=3.5296$) mark them as star-forming galaxies.
Such galaxies were shown to exhibit extended halos of metal-rich gas that cover a spatial region of at least 100 kpc, whose spatial and kinematic distributions may be identified by absorption lines that appear in the spectra of background objects \citep{fernandez2022,solimano2022,lopez2024,hernandez2026}.
The images of Source \#2 and Source \#5 are separated from each other in the image plane by less than $2''$, indicating that light from the background source (Source \#5) passes through the extended halo of the foreground source (Source \#2). This fortuitous alignment opens up a potentially interesting avenue toward mapping the distribution of the gas around the lensed foreground galaxy, Source \#2. To test whether this may be possible, we revisit the spectra for these objects extracted from the VLT/MUSE observations to search for absorption lines from Source \#2 in the spectra of the higher-redshift Source \#5. \autoref{fig:speccompare} shows the results of this inspection. We extract one region from the cube that covers the two Eastern merging images of Source \#2, and two regions covering the right and left images of Source \#5. The resulting spectra show pronounced iron and magnesium absorption features in the spectrum of Source \#2. The same features, i.e., iron and magnesium lines at the redshift of Source \#2, also appear in absorption in the spectrum of the background Source \#5. This indicates that further analysis of Source \#5 could trace the spatial and kinematic distribution of the gas halo around the foreground galaxy, Source \#2. We reconstruct the source plane at $z=1.972$ to calculate the impact parameter between the core of Source \#2, which is located at this redshift, and the core of Source \#5 as it appears at this redshift. We find that at this source plane, the separation between the blue core of Source \#2 and the two images of Source \#5 are 2.3 and 3.5 kpc.  Given the high magnification of Source \#2, and its location just outside of the peak epoch of star formation known as cosmic noon ($z\sim1-3$; see, e.g., \citealt{cosmicnoon}) this particular spatial arrangement may allow for a uniquely precise study of the circumgalactic medium around this galaxy \citep[c.f.][for such analyses of other galaxies]{lopez2018,lopez2020,berg2025}. The spectra extracted from the left and right components probe the halo at two projected radii, adding valuable spatial information about the physical properties of the extended halo.

\subsection{Inclusion of Light-Unaffiliated Mass Components} \label{sec:extshear}
The lens model presented in \citetalias{lagattuta2023} requires a non-negligible external shear component ($\gamma\sim0.05$), and also includes a DM halo around a `dark clump’, or LUMC, that is not anchored to any visible matter. Recent work by \cite{limousin2022} argued against incorporating any terms without sufficient physical motivation, as the necessity of these components in the lens model may be symptomatic of the limitations of parametric mass modeling. Given the new information we are able to add to the model with JWST imaging, we explored whether the external shear component and the dark clump are indeed necessary for the model. We define a component as necessary only if it produces an rms that is significantly lower ($\sim25\%$) than a model created without the component, or a BIC and AIC improvement greater than 2. 

 We inspect archival DESI Legacy Imaging Survey images from DR9, DR10, and DR11 \citep{desi2019} to search for any nearby large scale structures, galaxy clusters, or foreground galaxies that could be responsible for this term. We tentatively identify an overdensity of galaxies with similar colors to RXJ0437 (see~\autoref{sec:desi}), situated~1 Mpc to the North-East of the main cluster. Intriguingly, these objects lie in a direction that is compatible with the implied external shear component of the lens model, suggesting a possible physically motivated mass structure that could be responsible for the shear. We stress, however, that this structure has not been confirmed spectroscopically; additional study is required before we can confirm this possibility. 

To explore whether this overdensity of galaxies could be producing the residual shear observed in the \citetalias{lagattuta2023} model, we introduce a spatially constant shear term \citep[e.g.,][]{keeton1997} into the fiducial model. This term has two parameters: a magnitude ($\gamma$) and a position angle ($\theta_\gamma$). The use of this component in \lenstool~models typically adds flexibility that can result in a better statistical performance for the model. In the final shear model, we find $\gamma\sim0.07$ and $\theta_\gamma\sim23$. 
 
The direction of the shear term from this model is consistent with the direction of the shear from \citetalias{lagattuta2023}, and points toward the possible overdensity of galaxies noted above. However, the improvements to the fiducial model are not enough to justify this term's inclusion, as it is not strictly necessary to create a lens model that reproduces the observable constraints. The shear model obtains a mild improvement in the rms ($0\farcs20$, a $16\%$ difference) and yields BIC and AIC values that are only 1.8 times lower than the fiducial model. Since we lack strong evidence (e.g., spectroscopic confirmation of the galaxies that may be part of the overdensity) that associates the residual shear term with physical mass, and since the statistical improvements to the model are not very significant, we do not use shear in the fiducial model. We find that the large number of additional arcs we identify from the NIRCam imaging places extremely stringent limits on the inner mass distribution of the cluster. Quantitatively, we add 120 new constraints to the inner $\sim50$ kpc of the cluster through the identification of 17 new sub-clumps for Source \#1, 6 new sub-clumps for Source \#2, 10 new sub-clumps for Source \#5, Image \#12.4, and Sources \#15, 16, and 25. This abundance of information likely plays a significant role in forcing the model to assume a moderate ellipticity without requiring an external shear component.

However, we stress that we cannot rule out the inclusion of an external shear term completely. In this work, we judge its contribution to the model to add less value than the uncertainty we gain by including a component of mass that cannot be definitively associated with any light in the surrounding regions, since the statistical improvements it offers may also originate from the systematics inherent to parametric lens modeling \citep{limousin2022}. Future spectroscopic measurement of the possible overdensity of galaxies may change this assessment. 

On a smaller scale, we test the possible statistical improvements offered by the LUMC from \citetalias{lagattuta2023} in both the fiducial and shear models. We find that the benefit offered by this LUMC can be replicated and, in fact, improved on, by the inclusion of a group-scale halo on the third brightest cluster member in the south, which adds more additional mass in this region. We do not find sufficient evidence to suggest that a LUMC is necessary to reproduce the observed JWST constraints.

\subsection{Implications for $<10^9M\odot$ Substructure Detection}\label{sec:dmsubstructure}

\begin{figure*}
\centering
\begin{minipage}{1\linewidth}
    \hspace*{-7cm}\includegraphics[width=1.5\linewidth]{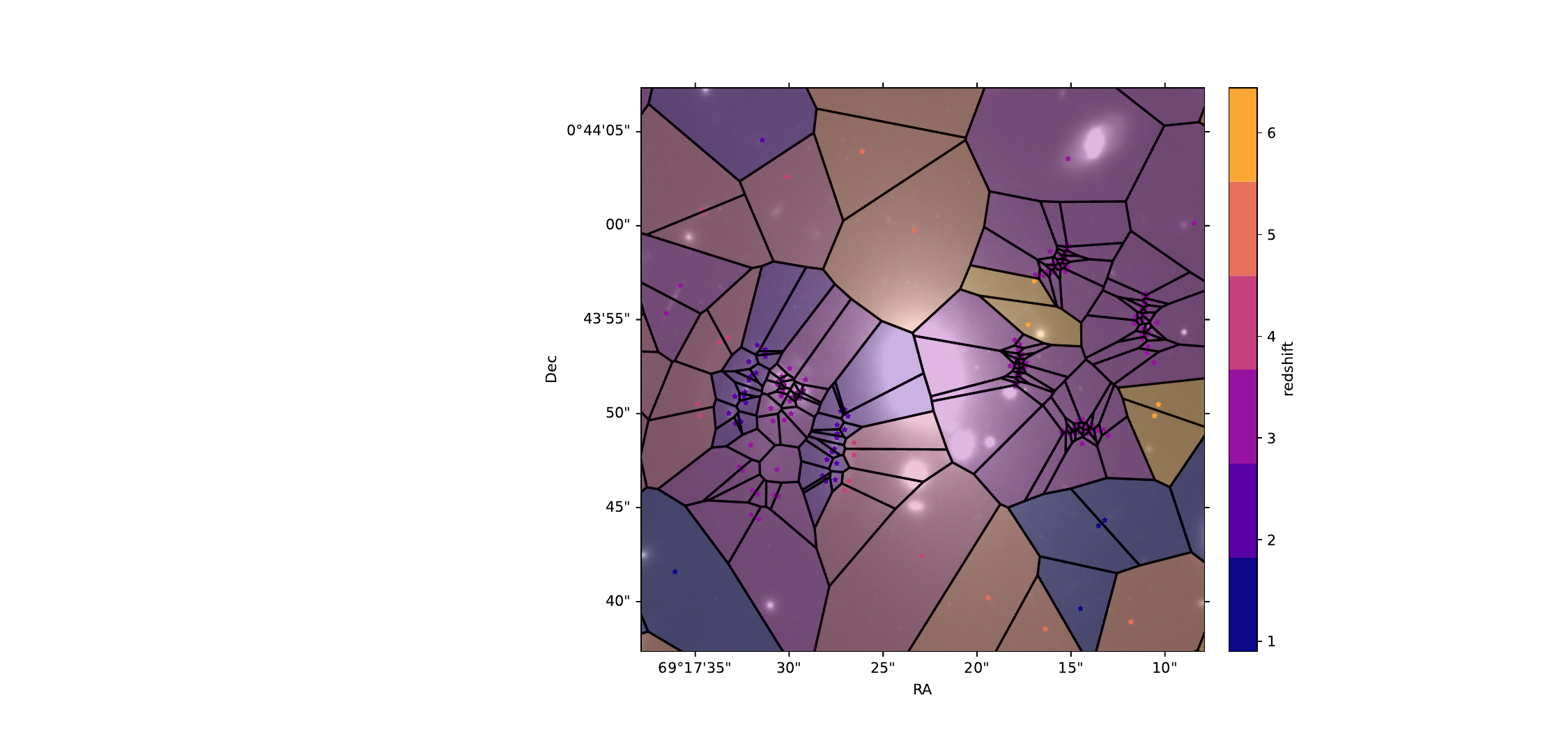}
\end{minipage}
    \caption{A $20\arcsec \times 20\arcsec$ cutout of the cluster centered around the BCG. The density of lensing constraints is shown by partitioning the spatial area into finite polygons whose size describes the distance between constraints, using a custom-coded Voronoi algorithm. The colorbar shows the spectroscopic and model-predicted redshifts of the arcs. The constraints are plotted as stars with colors corresponding to the values in the redshift colorbar. }
    \label{fig:spatialconstraints}
\end{figure*}

One of the primary purposes of this lens model is to serve as a global lensing potential solution for a forthcoming investigation of DM properties. Specifically, an accurate and precise lens model which probes the lensing potential with high spatial resolution is required in order to detect small-scale DM substructures. These low mass substructures ($M < 10^9 M_\odot$) are favored in $\Lambda$CDM, but are less present in alternative dark matter models, such as warm DM, self-interacting DM, or fuzzy dark matter descriptions of the Universe, since the subhalo mass function in these models is significantly depressed \citep{robertson2019,he2023}. Their detection or non-detection is thus a strong test of DM properties and of the cosmological makeup of the Universe as a whole. 

The number of arcs in RXJ0437, along with its three H-U systems, makes it a one-of-a-kind detection laboratory for these substructures. The presence of a lensing constraint at a spatial location places a tight restriction on the allowable mass in that region. If a lens model can appropriately model cluster mass and account for magnification effects on lensed images, and the baryonic mass components of the cluster (cluster members, ICL, etc) are appropriately accounted for, then dark matter substructures and fluctuations are able to be detected with high accuracy \citep{natarajan2017,he2023}. This measurement is especially sensitive to local perturbers; \citealt{he2025} found that a previously identified dark, highly concentrated subhalo in a lensed galaxy-galaxy system was more likely to be a dwarf satellite galaxy, though the possibility that the subhalo was entirely dark could not be ruled out. A combination of deep multi-band imaging, spectroscopy, and lens modeling that accounts for the mass and light of all observable components is crucial to ensure statistically sound results; these criteria are a driver of upcoming science missions like the Habitable Worlds Observatory, which will facilitate the detection of these substructures on scales of $10^8M\odot$ \citep{dressing2026}. Fortunately, in this work, we are already able to satisfy all three criteria to identify possible DM substructures in RXJ0437.

\citealt{lagattuta2026} has taken the first step toward the detection of these substructures using the HST-based model from \citetalias{lagattuta2023}. The authors use differences in the brightness and position of lensed clumps in the H-U systems to identify regions where substructure may be present. These H-U clumps should have the same relative position and luminosity in each of the four central multiple images, thanks to the intrinsic isotropic lensing properties of an H-U system. However, if there are significant variations in the brightness or position of the clumps relative to their appearance in other multiple images, then an additional mass component may be influencing the lensing of those clumps. This mass component is typically much smaller than the cluster halo, on order of $10^8-10^9M\odot$, and may represent a DM subhalo. 

A key requirement of this measurement is the need for a highly accurate lens model to correct for lensing magnification effects and constrain the position of the clumps in each multiple image.  \cite{lagattuta2026} identified a candidate DM subhalo of $2\times10^9M\odot$ near the second image of Source \#1 using the HST-based lens model, which is already a promising result. In future analysis, we will test this discovery using the updated JWST lens model we present in this work. We nearly triple the number of arcs used to constrain the mass of this cluster, and increase the amount of constraints in the inner $\sim50$ kpc region by 120. We also add 17 new sub-clumps to each multiple image of Source \#1, which adds higher fidelity to the measurement of positional differences of each clump between each multiple image.

Another useful feature of this work is that the model constraints exist at a variety of redshifts, ranging from $\sim0.9<z<6$. \autoref{fig:spatialconstraints} depicts this visually by showing the spatial density of constraints in the inner 20 arcsec$^2$ of the cluster, as well as showing the redshift distribution of these constraints. At the highest densities, we have a constraint every $\sim0\farcs01$, and on average in this region, we have a constraint every $\sim2\farcs5$. The constraints in the lens model achieve a tight spatial probe of this cluster in both 2D and 3D space. The smallest regions in this figure correspond to the regions with the highest density of arcs, which makes these locations prime areas for discovering DM substructures. We plan to explore these regions in an upcoming paper.

\subsection{Clumps Behind the BCG}
One possible avenue for further improving this model is by increasing the number of arcs in the region where the BCG is located, which is currently impeded by light from the BCG that obscures any multiple images that may be behind it in the NIRCam imaging. There are at least 6 predicted multiple images that lie within the region dominated by the BCG. We performed a basic subtraction of the BCG light using a Sersic profile, but were unable to conclusively identify any lensed images. Since their presence does not affect the mass distribution of the global lens model, their identification is outside the scope of this work. More refined modeling of the BCG light, as well as the nearby cluster member galaxies, may reveal the precise locations of these images, though since they are so demagnified they may not appear at all. 

\section{Summary and Future Work} \label{sec:conclusions}

In this work, we present an updated lens model of RXJ0437 based on new JWST NIRCam imaging, which increases the amount of multiple images used as constraints in the strong lensing model from 79 to \nconstraints. The large increase is due to \nsysnew new systems of lensed galaxies we identify from JWST, as well as from the resolved star-forming clumps in the two visible H-U systems and the highly-magnified galaxy lying close to the tangential and radial critical curves. The increase in constraints allows for a more complex parameterization of the model, allowing for a more physically-motivated mass distribution, but which increases the computational time needed for the model to reach convergence. In order to facilitate a statistically-meaningful exploration of the parameter space around the best-fit model, we used a progressive approach in which abridged, well-converged models with fewer constraints were used to identify and define priors on each optimized parameter that sampled the space around the best-fit model. This significantly reduces the computational time required for the final, fiducial model to converge and ensures that the parameter space is well-sampled in the area around the best-fit model.

In the model we present in this work, we achieve an image plane rms of \finalrms~using only mass components that are physically motivated by the presence of light. We achieve a high spatial density of constraints, with one constraint every $2.5$ square arcseconds on average in the inner $\sim50$ kpc of the cluster. We show that the use of an external shear component does not offer improvements to the model that are statistically meaningful enough to justify its use \citep{limousin2022}, which may be influenced by the high number of arcs in the inner part of the cluster. 

The lens model we present here has very tightly constrained magnifications (statistical uncertainties of a few \%) and positional uncertainties (variance of a few tenths of an arcsecond) along the critical curves near the H-U systems, which is beneficial for the detection of DM substructure. Magnifications allow for the accurate measurement of clump luminosities, which can be used to identify discrepancies in clump brightnesses between multiple images. These discrepancies can then be used to identify possible DM substructures, as shown in \cite{lagattuta2026}. In this work, we show that the mass within the areas around the H-U multiple images is well-constrained due to the abundance of arcs in these regions (one arc every $0.01$ square arcseconds). The area around Source \#2 also provides multiple sight-lines due to the high number of lensed systems at different redshifts that are concentrated in this area. This region may thus be of most immediate interest for searching for small DM substructures, which will be undertaken in future work.

These precise magnification measurements also lend themselves to studies of the lensed galaxies themselves. Evaluating spectroscopic line strengths and correcting for magnification effects for each of the lensed clumps in the most highly magnified galaxies can be used to gain understanding of clump ages, star formation history, and galaxy assembly history \citep{noguchi1999,bournard2014,sachdeva2017, khullar2026}. Clump sizes can also be measured to gain understanding of the local physics driving the formation and activity of these star-forming regions \citep{knutas2025,lapeer2026}. Future work can leverage spectroscopic measurements of the identifiable features in each of these clumps alongside the magnifications of each pixel in the lensed galaxies to understand the source properties of the galaxies on a clump-by-clump basis.

\begin{acknowledgments}

This work is based on observations made with the NASA/ESA/CSA James Webb Space Telescope. The data were obtained from the Mikulski Archive for Space Telescopes at the Space Telescope Science Institute, which is operated by the Association of Universities for Research in Astronomy, Inc., under NASA contract NAS 5-03127 for JWST. These observations are associated with program JWST-GO-06207 (Cycle 3).
Support for program JWST-GO-06207 was provided by NASA through grants from the Space Telescope Science Institute, which is operated by the Association of Universities for Research in Astronomy, Inc., under NASA contract NAS 5-03127. The imaging data described in this work can be obtained from the MAST archive at \dataset[10.17909/sn0v-w568]{\doi{10.17909/sn0v-w568}}.
Based on VLT/MUSE observations collected at the European Southern Observatory under ESO programmes 0104.A080, 106.21AD.001. 
MJ and BB acknowledge support from the United Kingdom Research and Innovation (UKRI) Future Leaders Fellowship `Using Cosmic Beasts to uncover the Nature of Dark Matter' (grant number MR/X006069/1).
ML acknowledges CNRS and CNES for support. 
AZ acknowledges support by the Israel Science Foundation Grant No. 864/23.
This research was supported in part through computational resources and services provided by Advanced Research Computing at the University of Michigan, Ann Arbor. 

\end{acknowledgments}

\facilities{HST, JWST (NIRCam), VLT (MUSE)}

\software{astropy \citep{2013A&A...558A..33A,2018AJ....156..123A,2022ApJ...935..167A},  
          Source Extractor \citep{1996A&AS..117..393B},
          \lenstool\ \citep{jullo07}, SAO ds9,
          Drizzlepac \citep{drizzlepac}}


\appendix

\section{VLT/MUSE Spectroscopy} \label{sec:newspecz}
In this Appendix, we show the spectra of the two new spectroscopic identifications that we make in this work. We also show the spectrum for the fourth image of Source \#12.

\begin{figure*}[htbp]
\begin{minipage}{1\linewidth}
\centering
    \includegraphics[width=0.6\linewidth]{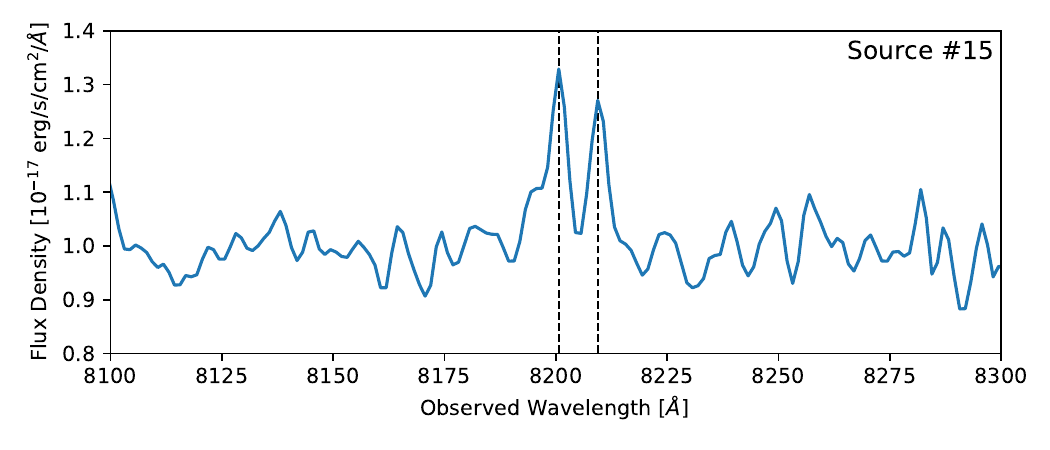}\\
    \includegraphics[width=0.6\linewidth]{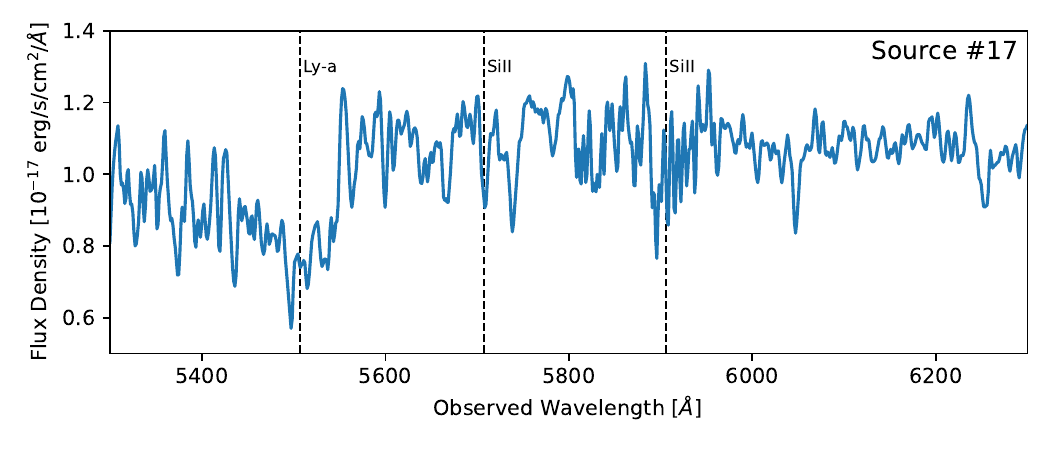}\\
    \includegraphics[width=0.6\linewidth]{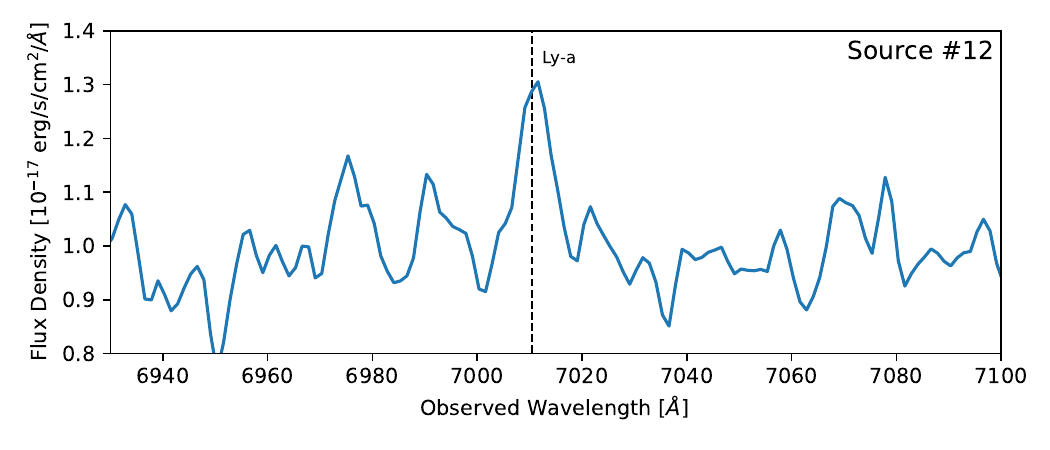} \\
    \includegraphics[width=0.15\linewidth]{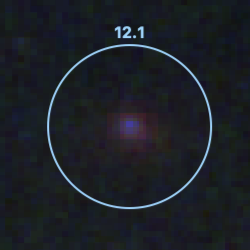}
    \includegraphics[width=0.15\linewidth]{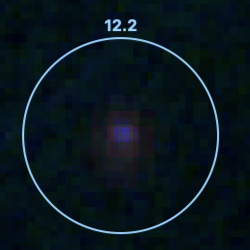}
    \includegraphics[width=0.15\linewidth]{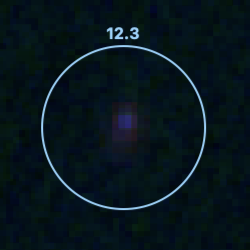}
    \includegraphics[width=0.15\linewidth]{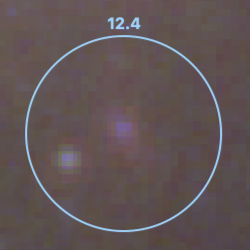}
\end{minipage}\\
    \caption{
    VLT/MUSE spectra for Sources \#12.4, \#15, and \#17. The redshift of Source \#12.4 is $z=4.7688$, consistent with the other three multiple images identified in \citet{lagattuta2023}. The redshift of Source \#15 is $3.301$. The redshift of Source \#17 is $3.523$. The four images of Source \#12 are shown in the bottom row; Image \#12.4 is a new identification in this work. The cyan circles show the approximate area used to extract the VLT/MUSE spectra for each image of Source \#12.
    }
    \label{fig:spectralextractions}
\end{figure*}

\newpage

\section{Clump Magnifications } \label{sec:clumpmag}
We measure the magnifications of the clumps in the two visible H-U systems (Source \#1 and Source \#2), and the most highly-magnified galaxy in the field (Source \#5). The magnifications are taken from the best-fit model. Errors are measured from a randomly selected sample of 100 models taken from the posterior distribution, and correspond to the 68\% confidence level. A truncated version of the table is shown here; we provide the full version in a separate machine-readable file attached to this work.

\startlongtable
\begin{deluxetable*}{cccc|ccccc} 
\tablecolumns{8} 
\tablecaption{Clump Magnifications} 
\tablehead{\colhead{Clump} &
            \colhead{R.A. [J2000]} & 
            \colhead{Dec. [J2000]} &
            \colhead{$\mu$} &
            \colhead{Clump} &
            \colhead{R.A. [J2000]} & 
            \colhead{Dec. [J2000]} &
            \colhead{$\mu$} &
            }    
\startdata 
101.1	&	69.28754	&	0.733044	&	$21.47_{-0.50}^{+0.37}$	&	108.4	&	69.288281	&	0.731143	&	$10.74_{-0.24}^{+0.29}$	\\
101.2	&	69.286404	&	0.732336	&	$23.25_{-1.06}^{+0.52}$	&	109.1	&	69.287658	&	0.732872	&	$20.75_{-0.44}^{+0.31}$	\\
101.3	&	69.287645	&	0.730273	&	$14.44_{-0.12}^{+0.58}$	&	109.2	&	69.286403	&	0.732003	&	$24.26_{-1.37}^{+0.23}$	\\
101.4	&	69.288328	&	0.730957	&	$6.42_{-0.28}^{+0.02}$	&	109.3	&	69.287413	&	0.73027	&	$15.14_{-0.02}^{+0.57}$	\\
102.1	&	69.287537	&	0.732996	&	$22.22_{-0.61}^{+0.32}$	&	109.4	&	69.28829	&	0.731198	&	$11.31_{-0.47}^{+0.06}$	\\
102.2	&	69.286427	&	0.732283	&	$24.46_{-1.37}^{+0.30}$	&	110.1	&	69.287608	&	0.732791	&	$23.63_{-0.44}^{+0.49}$	\\
102.3	&	69.287613	&	0.730297	&	$14.96_{-0.20}^{+0.51}$	&	110.2	&	69.286486	&	0.731989	&	$28.18_{-1.69}^{+0.30}$	\\
102.4	&	69.288308	&	0.730984	&	$7.55_{-0.19}^{+0.14}$	&	110.3	&	69.287408	&	0.730359	&	$17.23_{-0.18}^{+0.54}$	\\
103.1	&	69.287589	&	0.733036	&	$20.12_{-0.53}^{+0.24}$	&	110.4	&	69.288241	&	0.731206	&	$12.75_{-0.53}^{+0.09}$	\\
103.2	&	69.286373	&	0.732277	&	$22.24_{-1.06}^{+0.39}$	&	111.1	&	69.287558	&	0.732716	&	$27.08_{-0.78}^{+0.39}$	\\
103.3	&	69.287603	&	0.730249	&	$13.82_{-0.18}^{+0.44}$	&	111.2	&	69.286559	&	0.731984	&	$31.94_{-1.69}^{+0.70}$	\\
103.4	&	69.288334	&	0.73099	&	$7.12_{-0.31}^{+0.03}$	&	111.3	&	69.287399	&	0.730444	&	$20.03_{-0.08}^{+0.96}$	\\
104.1	&	69.287595	&	0.732998	&	$20.66_{-0.23}^{+0.54}$	&	111.4	&	69.288188	&	0.731237	&	$14.87_{-0.40}^{+0.36}$	\\
104.2	&	69.286386	&	0.732222	&	$22.83_{-1.10}^{+0.41}$	&	112.1	&	69.287658	&	0.732801	&	$21.87_{-0.62}^{+0.21}$	\\
104.3	&	69.287562	&	0.730249	&	$13.97_{-0.09}^{+0.51}$	&	112.2	&	69.28644	&	0.731953	&	$26.19_{-1.42}^{+0.37}$	\\
104.4	&	69.288321	&	0.731034	&	$8.26_{-0.31}^{+0.05}$	&	112.3	&	69.287375	&	0.730312	&	$16.24_{-0.02}^{+0.61}$	\\
104.5	&	69.297776	&	0.729263	&	$2.59_{-0.04}^{+0.02}$	&	112.4	&	69.288284	&	0.73124	&	$11.68_{-0.37}^{+0.18}$	\\
105.1	&	69.287573	&	0.732914	&	$22.42_{-0.48}^{+0.42}$	&	113.1	&	69.287627	&	0.732755	&	$23.43_{-0.72}^{+0.19}$	\\
105.2	&	69.286441	&	0.732161	&	$25.42_{-1.53}^{+0.20}$	&	113.2	&	69.28649	&	0.731935	&	$28.63_{-1.73}^{+0.29}$	\\
105.3	&	69.287518	&	0.730302	&	$15.41_{-0.02}^{+0.69}$	&	113.3	&	69.287366	&	0.730376	&	$17.90_{-0.15}^{+0.59}$	\\
105.4	&	69.288286	&	0.73108	&	$9.94_{-0.45}^{+0.01}$	&	113.4	&	69.28824	&	0.731249	&	$13.33_{-0.59}^{+0.05}$	\\
105.5	&	69.297794	&	0.72924	&	$2.58_{-0.04}^{+0.02}$	&	113.5	&	69.297815	&	0.729165	&	$2.57_{-0.04}^{+0.02}$	\\
106.1	&	69.287576	&	0.732863	&	$23.37_{-0.44}^{+0.50}$	&	114.1	&	69.287817	&	0.732952	&	$16.61_{-0.19}^{+0.32}$	\\
106.2	&	69.286466	&	0.732108	&	$26.39_{-1.41}^{+0.43}$	&	114.2	&	69.286234	&	0.731894	&	$19.44_{-0.90}^{+0.24}$	\\
106.3	&	69.287481	&	0.730329	&	$16.18_{-0.05}^{+0.65}$	&	114.3	&	69.28734	&	0.730108	&	$12.58_{-0.01}^{+0.43}$	\\
106.4	&	69.28826	&	0.731116	&	$10.89_{-0.30}^{+0.24}$	&	114.4	&	69.288405	&	0.731258	&	$9.04_{-0.35}^{+0.05}$	\\
106.5	&	69.297804	&	0.729219	&	$2.58_{-0.04}^{+0.02}$	&	115.1	&	69.28759	&	0.732646	&	$27.62_{-0.78}^{+0.39}$	\\
107.1	&	69.287629	&	0.73291	&	$20.75_{-0.55}^{+0.21}$	&	115.2	&	69.28657	&	0.73188	&	$33.57_{-1.60}^{+0.95}$	\\
107.2	&	69.286402	&	0.732093	&	$23.88_{-1.34}^{+0.24}$	&	115.3	&	69.287334	&	0.730469	&	$21.51_{-0.03}^{+0.90}$	\\
107.3	&	69.287473	&	0.730268	&	$14.59_{-0.09}^{+0.50}$	&	115.4	&	69.288168	&	0.731304	&	$16.60_{-0.39}^{+0.49}$	\\
107.4	&	69.288306	&	0.731125	&	$9.98_{-0.35}^{+0.12}$	&	116.1	&	69.287702	&	0.732731	&	$21.99_{-0.50}^{+0.30}$	\\
107.5	&	69.297786	&	0.729219	&	$2.59_{-0.04}^{+0.02}$	&	116.2	&	69.286445	&	0.731807	&	$27.36_{-1.64}^{+0.23}$\\ 
108.1	&	69.287615	&	0.732881	&	$21.65_{-0.41}^{+0.40}$	&	116.3	&	69.287282	&	0.730344	&	$17.82_{-0.04}^{+0.65}$\\
108.2	&	69.286432	&	0.732079	&	$25.05_{-1.22}^{+0.49}$	&	116.4	&	69.288254	&	0.73133	&	$13.69_{-0.51}^{+0.16}$	\\
108.3	&	69.287463	&	0.730293	&	$15.29_{-0.12}^{+0.51}$	&	116.5	&	69.297806	&	0.729134	&	$2.57_{-0.04}^{+0.02}$\\
...	&	...	&	...	&	...	&	...	&	...	&	...	& ...\\
\enddata 
 \tablecomments{Magnifications of each clump in Sources \#1, \#2, and \#5. The full table is available in the electronic version of this manuscript.}
\end{deluxetable*}

\section{Time Delay Measurements}\label{sec.tdelay}

Multiple images of transients are an effective means of placing constraints on $H_0$, as the time lag between the appearance of the transient in each image is a direct measure of the light travel time. Given the high magnification of the background source galaxies in the main H-U systems, we elect to provide time delay measurements for each image in the event that a transient source is ever identified in this cluster. If a transient appears in any of the H-U systems, then the expected reappearance in any of the four central images is on the order of one to three weeks, making the prior calculation of these values useful for follow-up observational programs.

\startlongtable
\begin{deluxetable}{c|clllllllll} 
\tablecolumns{10} 
\tablecaption{Predicted time delays from fiducial model} 
\tabletypesize{\small}
\tablehead{\colhead{Source} & 
            \colhead{Clump} &
            \colhead{$\Delta\tau_{\mathrm{2-1}}$ } & 
             \colhead{$\sigma_{\mathrm{2-1}}$ } & 
            \colhead{$\Delta\tau_{\mathrm{3-1}}$ } & 
             \colhead{$\sigma_{\mathrm{3-1}}$ } & 
            \colhead{$\Delta\tau_{\mathrm{4-1}}$ } & 
             \colhead{$\sigma_{\mathrm{4-1}}$ } & 
            \colhead{$\Delta\tau_{\mathrm{5-1}}$ } & 
             \colhead{$\sigma_{\mathrm{5-1}}$ } & \\ \colhead{} & \colhead{} & \colhead{days} & \colhead{days} & \colhead{days} & \colhead{days} &\colhead{days} &\colhead{days} &\colhead{days} &\colhead{days} 
            }    
\startdata 
Fiducial Model & 1 & -24.6 & 0.3 & 117.1 & 1.8 & 137.1 & 1.5 & \nodata & \nodata \\
Source \#1&2 & -23.4 & 0.3 & 103.9 & 1.7 & 124.3 & 1.4 & \nodata & \nodata \\
$z=2.9732$&3 & -30.4 & 0.4 & 103.1 & 1.8 & 128.0 & 1.4 & \nodata & \nodata \\
&4 & -30.5 & 0.4 & 90.9 & 1.7 & 117.0 & 1.3 & -8861.2 & 34.0 \\
&5 & -26.1 & 0.3 & 74.6 & 1.5 & 99.2 & 1.2 & -8967.2 & 34.2 \\
&6 & -25.4 & 0.3 & 62.2 & 1.4 & 87.4 & 1.1 & -9026.3 & 34.3 \\
&7 & -33.3 & 0.4 & 61.1 & 1.5 & 91.9 & 1.1 & -8958.0 & 34.2 \\
&8 & -30.8 & 0.4 & 57.8 & 1.5 & 87.1 & 1.1 & \nodata & \nodata \\
&9 & -36.7 & 0.4 & 45.9 & 1.4 & 80.5 & 1.0 & \nodata & \nodata \\
&10 & -27.9 & 0.3 & 40.9 & 1.3 & 69.7 & 0.9 & \nodata & \nodata \\
&11 & -20.6 & 0.3 & 37.5 & 1.1 & 61.0 & 0.8 & \nodata & \nodata \\
&12 & -34.9 & 0.4 & 32.1 & 1.3 & 66.8 & 0.9 & \nodata & \nodata \\
&13 & -29.5 & 0.3 & 30.5 & 1.2 & 61.4 & 0.8 & -9134.2 & 34.6 \\
&14 & -69.5 & 0.7 & 22.8 & 1.6 & 81.7 & 1.0 & \nodata & \nodata \\
&15 & -22.7 & 0.3 & 20.6 & 1.0 & 47.1 & 0.7 & \nodata & \nodata \\
&16 & -40.0 & 0.4 & 10.7 & 1.1 & 51.3 & 0.7 & -9131.6 & 34.5 \\
&17 & -47.4 & 0.5 & 8.7 & 1.2 & 54.9 & 0.8 & -9070.8 & 34.3 \\
&18 & -56.5 & 0.6 & -3.6 & 1.2 & 50.5 & 0.7 & -9051.3 & 34.2 \\
&19 & -55.7 & 0.6 & -9.3 & 1.1 & 44.9 & 0.7 & -9088.2 & 34.4 \\
&20 & -43.5 & 0.5 & -7.4 & 1.0 & 37.8 & 0.6 & -9196.3 & 34.6 \\
&21 & -47.6 & 0.5 & -15.2 & 1.0 & 33.9 & 0.5 & -9200.9 & 34.6 \\
&22 & -67.6 & 0.6 & -35.2 & 1.0 & 31.2 & 0.5 & -9121.1 & 34.3 \\
&23 & -79.1 & 0.7 & -55.2 & 1.0 & 22.4 & 0.4 & -9134.0 & 34.2 \\
&24 & -114.0 & 0.9 & -88.7 & 1.1 & 19.0 & 0.3 & \nodata & \nodata \\
\hline
Fiducial Model & 1 & 80.9 & 0.8 & 71.9 & 1.0 & -14.9 & 0.3 & \nodata & \nodata \\
Source \#2 & 2 & 53.9 & 0.6 & 38.7 & 0.8 & -18.8 & 0.4 & -7786.8 & 33.9 \\
$z=1.972$ & 3 & 49.3 & 0.5 & 35.6 & 0.8 & -15.8 & 0.3 & -7831.2 & 34.1 \\
 & 4 & 43.9 & 0.5 & 31.8 & 0.7 & -13.4 & 0.3 & -7878.8 & 34.2 \\
 & 5 & 77.7 & 0.8 & 76.4 & 0.8 & -2.5 & 0.1 & \nodata & \nodata \\
 & 6 & 44.2 & 0.5 & 41.7 & 0.6 & -1.8 & 0.1 & \nodata & \nodata \\
 & 7 & 52.4 & 0.6 & 51.5 & 0.6 & -0.8 & $<0.1$ & \nodata & \nodata \\
 & 8 & 52.5 & 0.6 & 52.0 & 0.6 & -0.4 & $<0.1$ & \nodata & \nodata \\
 \hline 
 Fiducial Model & 1 & -0.9 & $<0.1$ & -9851.4 & 35.5& \nodata & \nodata& \nodata & \nodata \\
 Source \#5 & 2 & -1.7 & $<0.1$ & -9999.0 & 35.9& \nodata & \nodata& \nodata & \nodata \\
 $z=3.5296$& 3 & -1.3 & $<0.1$ & -10023.9 & 36.0& \nodata & \nodata& \nodata & \nodata \\
 & 4 & -0.7 & $<0.1$ & -10086.0 & 36.2& \nodata & \nodata& \nodata & \nodata \\
 & 5 & -1.8 & $<0.1$ & -10074.0 & 36.1& \nodata & \nodata& \nodata & \nodata \\
 & 6 & -0.8 & $<0.1$ & -10123.3 & 36.3& \nodata & \nodata& \nodata & \nodata \\
 & 7 & -1.9 & $<0.1$ & -10111.4 & 36.3& \nodata & \nodata& \nodata & \nodata \\
 & 8 & -0.6 & $<0.1$ & -10150.4 & 36.4& \nodata & \nodata& \nodata & \nodata \\
 & 9 & -0.4 & $<0.1$ & -10203.4 & 36.5 & \nodata & \nodata& \nodata & \nodata\\
 & 10 & -1.5 & $<0.1$ & -10258.0 & 36.7& \nodata & \nodata& \nodata & \nodata \\
 & 11 & -0.7 & $<0.1$ & -10299.4 & 36.7& \nodata & \nodata& \nodata & \nodata \\
 \hline
\enddata 
\tablecomments{Predicted time delays, calculated from the best-fit fiducial model, for the two visible H-U systems (Source \#1 and Source \#2), as well as for the highly magnified central image (Source \#5). The arrival time $\Delta\tau$ for each image of each clump is shown relative to image 1, and $\sigma$ shows the standard deviation of this measurement. 
}\label{tab.fiducialtdelay}
\end{deluxetable}

\section{A Wide-Field View of RXJ0437 } \label{sec:desi}
The figure included in this appendix shows an expanded view of this cluster to highlight the location of the possible overdensity of galaxies discussed in~\autoref{sec:extshear}.

\begin{figure*}[htbp]
 \centering
    \includegraphics[width=1\linewidth]{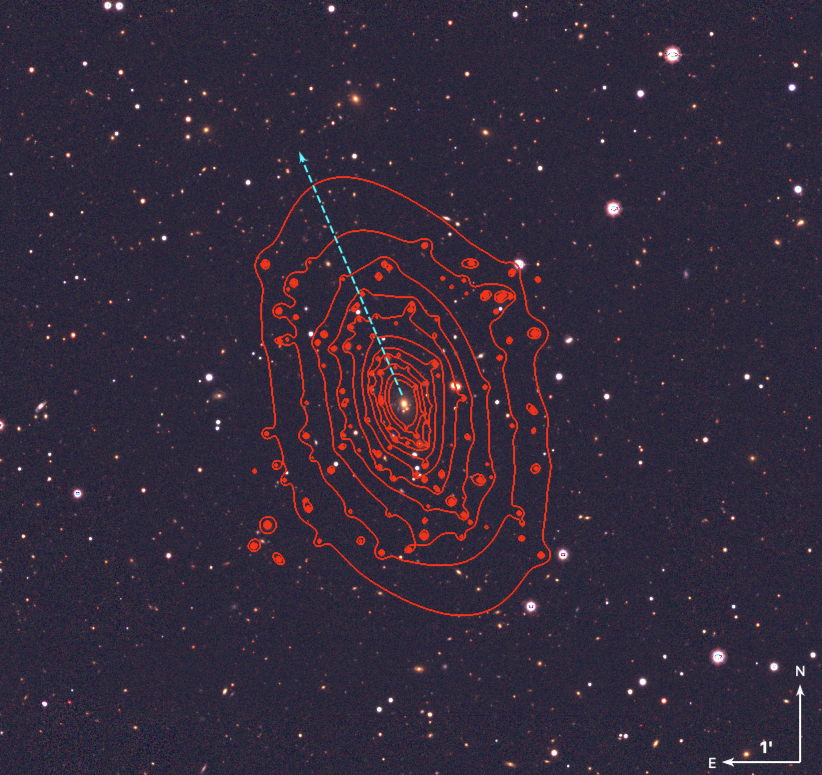}
    \caption{Mass contours from the best-fit lens model are overlaid in red on top of a 11'x11' cutout from DECam imaging. The cyan dashed line indicates the direction of the residual shear component obtained from the lens model. A possible overdensity of galaxies is located just to the north of the end of this arrow. 
    }\label{fig.desicutout}
\end{figure*}

\newpage


\bibliography{bib}{}
\bibliographystyle{aasjournal}



\end{document}